%% file: london.tex
\documentstyle[icproc,psfig]{article}

\begin{document}


\def\epsfig{\psfig}
\newcommand{\ra}{\rightarrow}
\newcommand{\ko}{K^0}
\newcommand{\be}{\begin{equation}}
\newcommand{\ee}{\end{equation}}
\newcommand{\bea}{\begin{eqnarray}}
\newcommand{\eea}{\end{eqnarray}}
%
\def\D{D}
\def\Jnote#1{{\bf[JL: #1]}}
\def\Snote#1{{\bf[SF: #1]}}
\def\addref#1{[{\bf add ref} #1] }
\def\cto{$\!\!^{-}$}
%
\lecture{DUALITY IN STRING THEORY}

\author{STEFAN F\"ORSTE
}

\address{Sektion Physik, Universit\"at M\"unchen, 
Theresienstra\ss e 37\\ D-80333 M\"unchen, Germany}

\author{JAN LOUIS
\thanks{Lecture given by J.~Louis at the Workshop on Gauge Theories,
Applied Supersymmetry and Quantum  Gravity,
Imperial College, London, UK, July 5--10,  1996.}
}

\address{Fachbereich Physik, Martin-Luther-Universit\"at,\\
D-06099 Halle, Germany}


\maketitle 
\abstracts{
In this lecture we review  some of the recent
developments in string theory on an introductory and 
qualitative level.
In particular we focus on $S-T-U$ dualities of toroidally
compactified ten-dimensional string theories and 
outline the connection to M-theory. 
Dualities among string vacua 
with less supersymmetries in  six
and four space-time dimensions is
 discussed and the concept of F-theory
is briefly presented.}

\section{Introduction}
During the past two years string theory has seen 
spectacular progress;
for the first time it has been possible
to control a subset of the interaction in the strong
coupling regime of string theory.
This is due to the observation that many (if not all)
of the perturbatively distinct string theories
are related when all quantum corrections are taken into 
account.\cite{FILQ}$\!\!\!^{-}$\cite{gabriel3}
In particular it has been observed that 
often the strong coupling regime of one string theory
can be mapped to the weak coupling regime of another,
perturbatively different string theory.
This situation is termed duality among string theories
and it offers the compelling picture 
that the  known perturbative string theories
are merely different regions in  the moduli space
of one underlying theory termed `M-theory'. The 
purpose of this lecture is to review on an 
introductory level some of the recent advances in 
establishing such string dualities.\footnote{We 
omit most
of the technical details and refer the reader to the
original literature. 
There are a number of nice review
lectures \cite{schwarzrev}$\!\!^{-}$\cite{townsendrev}
which also include a more extended list of references.}
However, this lecture does not  cover all
aspects of string duality but rather is meant to
compliment the
lectures by C.~Bachas, M.~Duff, A.~Giveon and
F.~Quevedo at this meeting.

This lecture is organized as follows.
In section~2 we discuss the `old'
perturbative string theory.
In sections~3--5 we review  T--, S-- and U--dualities 
of toroidally compactified
string theories. Section~6
introduces M-theory and its connection to 
ten-dimensional
string theory. In sections~7 and 8  we discuss
string dualities for compactifications 
with six (section~7) and four (section~8)
space-time dimensions
where some of the 
supercharges are broken. 

%
\section{Perturbative string theory }
\label{sec:prod}
\begin{figure}[h]
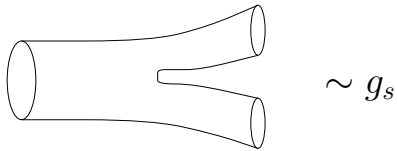
  
\begin{center}
\input fig1.pstex_t
\end{center}
\caption{Fundamental string interaction}
\label{fig:fig1}
\end{figure}

\begin{figure}
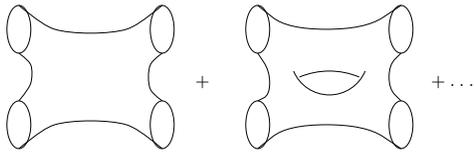
  
\begin{center}
\input fig2.pstex_t
\end{center}
\caption{String perturbation theory. 
The order of $g_s$ in the perturbative 
expansion is governed by the number of handles in the world sheet.
}
\label{fig:fig2}
\end{figure}

In string theory the fundamental objects are one-dimensional strings
which, as they move in time, sweep out a two-dimensional worldsheet
$\Sigma$.
This worldsheet is embedded in some higher dimensional target space
which is identified with a Minkowskian space-time.
Particles in this target space  appear as (massless) 
eigenmodes of the  string  and their scattering amplitudes are 
generalized by appropriate  scattering amplitudes of strings.
Strings can be open or closed, oriented or unoriented.\footnote{For an introduction to string theory 
we refer to the literature.\cite{tex1}$\!\!^{-}$\cite{tex4}}
String scattering amplitudes are built from
a  fundamental vertex;
for closed strings
this vertex is depicted in figure
\ref{fig:fig1}.\footnote{Open string scattering
is discussed in the lecture by C.~Bachas.}
It represents the splitting of a string or 
the joining of two strings and 
the strength of this interaction is governed by 
a  dimensionless string coupling constant $g_s$.
Out of the fundamental
vertex one composes all possible 
closed string scattering
amplitudes ${\cal A}$, for example the four-point amplitude
shown  in figure \ref{fig:fig2}.
The expansion in the topology of the Riemann surface
(i.e.~the number of holes in the surface)
coincides with  a power series expansion in the string
coupling constant formally written as
\begin{equation} \label{eq:pert}
{\cal A}\, =\, \sum_n g_s^{2n+2}\, {\cal A}^{(n)}\ ,
\end{equation}
where ${\cal A}^{(n)}$ is the scattering amplitude
on a Riemann surface of genus $n$.
The exponent $2n+2$ of $g_s$ is nothing but (minus) the Euler characteristic
of a two dimensional surface with $n$ handles and four boundaries
corresponding to the two incoming and the two outgoing strings.
In all string theories there is a massless scalar field
$\D$ called the dilaton which couples to 
the Gauss-Bonnet density of the world sheet.
Therefore the vacuum expectation value (VEV) of the dilaton determines the size of the string coupling and 
one finds \cite{AAT,tex1}
\begin{equation}
g_s\, =\, e^{\langle\D\rangle}\ .
\end{equation}
$g_s$ is a free parameter since the dilaton is a flat 
direction (a modulus)
of the effective potential.
String perturbation theory is 
restricted to the region of parameter space
(which is also called the moduli space) where 
$g_s < 1$ and the tree level amplitude
(genus $0$)
is the dominant contribution with higher 
loop amplitudes suppressed by higher powers 
of $g_s$.

The interactions of the string are governed by a two-dimensional 
field theory on the world-sheet $\Sigma$.
${\cal A}$ can be interpreted as an unitary scattering amplitude 
in the target space
whenever this two-dimensional field theory is
conformally invariant.
This condition puts a restriction on the number
of space-time dimensions $d$ and the space-time spectrum.
The known consistent string theories 
necessarily have $d \le 10$ and they
are particularly simple in 
the maximal possible dimension $d=10$.\footnote{An 
additional constraint arises from the requirement of 
modular invariance
of one-loop amplitudes which results in an 
anomaly free spectrum of the
corresponding ten-dimensional low energy effective 
theory.\cite{ASNW}}
There are only five consistent string theories
in $d=10$: the type IIA, the type IIB, the heterotic 
$E_8\times E_8$, the heterotic $SO(32)$
and the type I $SO(32)$ string;
their massless spectra are  summarized
in Table \ref{tab:tab2}.
\begin{table}[t]
\caption{Consistent string theories in ten dimensions.} \label{tab:tab2}
\begin{center}\begin{tabular}[h]{|l|l|l|l|l|}
\hline 
   & \# of $Q$'s& \# of $\psi_\mu$'s & \multicolumn{2}{c|}{bosonic spectrum}\\ 
\hline
IIA & 32 & 2& NS-NS & $g_{\mu\nu}$, $b_{\mu\nu}$, $\D$ \\ \cline{4-5}
 & & 
& R-R & $A_{\mu}$, $C_{\mu\nu\rho}$
\\ \hline\hline
IIB & 32 & 2& NS-NS &  $g_{\mu\nu}$, $b_{\mu\nu}$, $\D$ \\ \cline{4-5}
 & & 
& R-R & 
$c^{*}_{\mu\nu\rho\sigma}$, $b_{\mu\nu}^{\prime}$, $\D^{\prime}$ \\
\hline\hline
heterotic & 16 & 1 &\multicolumn{2}{c|}{$g_{\mu\nu}$, $b_{\mu\nu}$, $\D$}\\
$E_8\times E_8$ & & &\multicolumn{2}{c|}{$A_{\mu}^a$ in
 adjoint of $E_8\times E_8$ }
\\ \hline\hline
heterotic & 16 & 1 &\multicolumn{2}{c|}{$g_{\mu\nu}$, $b_{\mu\nu}$, $\D$}\\
$SO(32)$ & & &\multicolumn{2}{c|}{$A_{\mu}^a$ in
 adjoint of $SO(32)$ }
\\ \hline\hline
type I & 16 & 1 &NS-NS  & $g_{\mu\nu}$,  $\D$\\ \cline{4-5}
$SO(32)$ & & &open string &$A_{\mu}^a$ in
adjoint of $SO(32)$ \\ \cline{4-5}
 & & &R-R & $b_{\mu\nu}$
\\ \hline\end{tabular}\end{center} 
\end{table}
All five theories are space-time supersymmetric.
The type II theories have 32 supercharges $Q$
and  two gravitinos $\psi_\mu$;
in type IIA they have opposite chirality while in 
type IIB they have the same chirality.
This is often referred to as the non-chiral and  chiral 
$N=2$ supersymmetry in $d=10$.
The other three string theories have only half 
of the supercharges (16) or what is called
$N=1$ supersymmetry.\footnote{In space-time dimensions
other than 10 there are ambiguous definitions
of what is meant by $N$. Therefore
it is convenient to always just count the total number of 
supercharges and this is what we always do 
in this lecture.}
The heterotic string theories  have 
non-Abelian gauge symmetries and thus  vector bosons
and their superpartners  (gauginos) in the adjoint
representation of the gauge group.
In addition to the closed string theories 
there is one supersymmetric 
consistent theory (called type I)
containing unoriented open and closed strings with $SO(32)$
Chan-Paton factors coupling to the ends 
of the open string.

The bosonic spectrum of type I and type II theories
can appear in two distinct
sectors (NS-NS or R-R) 
depending on the boundary conditions 
of the worldsheet fermions.
Also note that 
the metric $g_{\mu\nu}$,
the antisymmetric tensor $b_{\mu\nu}$ 
and the dilaton $\D$ are common
to all five vacua. 

\section{T--duality}
The  ten-dimensional string theories can be compactified
to obtain theories with a lower number of space-time
dimensions $d$. The
simplest of such compactifications are
toroidal compactifications where the internal manifold
is a $n$-dimensional torus $T^{n}$ ($n=10-d$).
Such compactifications leave all supercharges unbroken.

For simplicity we start by considering closed string
theories with 
one compact dimension (or a $S^1$-compactification).
In this case there are
nine space-time coordinates $X^\mu$
satisfying the boundary conditions\footnote{The 
indices $\mu,\nu$ always denote
the space-time directions, 
i.e.~$\mu = 0,\ldots,d-1$.}
\begin{equation}
X^\mu \left(\sigma = 2\pi , \tau\right) 
= X^\mu\left(\sigma = 0 , \tau \right),
\end{equation}
and one internal coordinate $Y$ 
which can wrap  $m$ times around the $S^1$ of radius $R$,
\begin{equation} \label{eq:compact}
Y\left( \sigma = 2\pi , \tau \right) 
= Y\left( \sigma = 0, \tau\right) +2\pi m R\ .
\end{equation}
The massless spectrum of the nine-dimensional theory 
includes the two Abelian Kaluza--Klein gauge bosons
$g_{\mu 10}$ and $b_{\mu 10}$ leading to a 
gauge group $G=U(1)^2$.
In addition there is a massless 
scalar field $g_{10\, 10}$ 
which is a flat direction of the effective potential
and which is directly related to the radius $R$. 
The appearance of flat directions is a 
generic feature of string compactifications
and such scalar fields are called moduli.
For the case at hand this moduli space is 
one-dimensional and hence there is a 
one parameter family of inequivalent 
string vacua.\footnote{The different solutions
of a given string theory are often referred to as 
the vacuum states of that string theory
or simply as the string vacua.}
The boundary condition 
(\ref{eq:compact}) leads to a quantization of the 
internal  momentum component $p_{10}$
and a whole tower of massive Kaluza--Klein
states labelled by an integer $k$.
In addition there are also massive winding modes 
labelled by $m$ and altogether one finds
\begin{equation}\label{eq:mass}
 M^2 =  \frac{k^2}{R^2} + \frac{m^2 R^2}{4} +\,
(N+\tilde N-2)\ ,
\end{equation}  
where $N$ and $\tilde N$ are the number operators
of the left and right moving oscillator excitations.
This spectrum is invariant under the exchange of 
$R$ with
$2/R$ if simultaneously the winding number 
$m$ is exchanged with the momentum
number $k$.
This is the first duality -- 
a $Z_2$ invariance of the spectrum,
called {\sl T--duality}.\footnote{ 
For an extensive review about T--duality 
including further references we refer the reader to the
literature.\cite{j-gang}}

This simple example already displays another 
generic feature of string compactifications --
special points (or subspaces)
in the moduli space where the gauge symmetry is enhanced. 
For fixed $R=\sqrt{2}$ there are four additional
massless gauge bosons corresponding to  
$mk = \pm 1, N +\tilde N = 1$. 
These states combine with the
two $U(1)$ gauge fields and enlarge the 
$U(1)^2$ gauge symmetry to\footnote{Note 
that the enhancement of the gauge symmetry 
does not change the rank of $G$ but only its size.}
\begin{equation}
U(1)\times U(1) \longrightarrow SU(2)\times SU(2).
\end{equation} 

The exact same generic features -- 
Abelian Kaluza--Klein gauge bosons
and a moduli space with non-Abelian enhancement on special
subspaces -- also occur for higher dimensional toroidal
compactifications on a torus $T^n$.
The massless states at a
generic point in the moduli space include 
the Kaluza--Klein gauge bosons
$g_{\mu i}$ and $b_{\mu i}$  of the gauge
group $G=U(1)^{2n}$ and the toroidal moduli
$g_{ij}$, $b_{ij}$ parameterizing a moduli space of 
inequivalent string vacua.\footnote{
The indices $i,j$ run over the internal dimensions,
i.e.~$i,j=1,\ldots,n=10-d$.}
This moduli space is found to be the $n^2$-dimensional
coset space
\cite{narain}
\begin{equation}\label{eq:Mtoroidal}
{\cal M}=\left. \frac{SO(n,n)}{SO(n)\times SO(n)}\right/\Gamma_T \ ,
\end{equation}
where 
\begin{equation}\label{eq:Ttoroidal}
\Gamma_T = SO\left(n,n, Z\right)
\end{equation}
is the T--duality group relating equivalent string vacua.

In heterotic or type I theories there are $n$
additional scalars $A_{i}^a$
transforming in the adjoint representation
of  $E_8 \times E_8$ 
or $SO(32)$. 
However, only the $16\cdot n$ 
scalars in the Cartan subalgebra 
are flat directions and their (generic) VEVs
break the non-Abelian gauge symmetry
to $U(1)^{16}$.
Together with the toroidal 
 moduli $g_{ij}$, $b_{ij}$ 
they parameterize the $n(n+16)$ dimensional
moduli space \cite{narain}
\begin{equation}\label{eq:Mhettoroidal}
{\cal M}=\left. \frac{SO(n,n+16)}{SO(n)\times SO(n+16)}\right/\Gamma_T ,
\end{equation}
with the T--duality group
\begin{equation}\label{eq:Thettoroidal}
\Gamma_T = SO\left(n,n+16, Z\right)\ .
\end{equation}

Thus, in space-time dimensions less than ten there is a 
moduli space of inequivalent string vacua and at generic
points in this moduli space
the gauge group is 
$G=U(1)^{2n+16}$. 
On special
subspaces of the moduli space 
there can be non-Abelian enhancement of the 
$U(1)^{16}$ factor,  at most up to 
the original $E_8 \times E_8$ 
or $SO(32)$.
Furthermore, a discrete T--duality group identifies 
(equivalent) string vacua. 

It has also been
shown that below ten dimensions
the heterotic $E_8 \times E_8$ theory and
the heterotic $SO(32)$ theory are continuously connected
in the moduli space. That is, the two theories 
sit at different points of the
same moduli space of one and the same heterotic 
string theory.\cite{gins}
A very similar situation is found for type II theories 
below ten dimensions.
Although the massless spectrum of type IIA and type IIB
looks rather different in $d=10$ it precisely matches
 in 
$d\le 9$ as can be seen from  table \ref{tab:spectra}.
\begin{table}[t]
\caption{Massless type II fields in nine dimensions}\label{tab:spectra}
\begin{center}\begin{tabular}{|l|l|l|}
\hline & NS-NS & R-R 
\\ \hline\hline
IIA & $g_{\mu\nu}$, $b_{\mu\nu}$, $\D$, $g_{\mu 10}$, $b_{\mu 10}$ &
$A_\mu$, $A_{10}$, $C_{\mu\nu\rho}$, $C_{\mu\nu 10}$ \\ \hline
IIB & $g_{\mu\nu}$, $b_{\mu\nu}$, $\D$, $g_{\mu 10}$, $b_{\mu 10}$ &
$b_{\mu 10}^{\prime}$, $\D^{\prime}$, $C_{\mu\nu\rho 10}$, 
$b^{\prime}_{\mu\nu}$ \\ \hline 
\end{tabular}\end{center}\end{table}
Their low energy effective actions can
also be shown to agree \cite{berg} and furthermore,
a careful analysis
of the limits $R\to 0$ and $R\to\infty$ reveals 
a flip of the chiralities of the space-time
fermions as is necessary
to connect the non-chiral type IIA with the chiral
Type IIB theory.\cite{DHS,d-brane1}
Thus, for $d\le 9$ type IIA and type IIB
also sit in the same moduli space and 
are T-dual to each other in that 
type IIA at a large compactification radius 
is equivalent to
type IIB at a small compactification radius 
and vice versa.

Finally, toroidally compactified 
type I theories naively do not have a T--duality
symmetry.
However, once  extended objects -- termed 
D-branes -- are included as possible configurations
type I theories also 
are T-dual.\cite{d-brane1}$\!\!^{-}$\cite{d-brane10} 
This exciting aspect of the recent developments is 
covered  in the lecture of C.~Bachas 
and elsewhere \cite{d-brane4}
and will not be
discussed any further here.

%
\section{S--duality}
S--duality refers to the quantum equivalence of two theories 
$A$ and $B$
which are perturbatively distinct.
Generically the strong coupling regime of $A$ is mapped
to the weak coupling regime of $B$ and 
simultaneously
the perturbative excitations of $A$
are mapped to the non-perturbative excitations of the dual theory $B$
and vice versa.
Such a relation is of prime interest since it opens up
the possibility to control the strong coupling regime of 
both theories.\footnote{The T--duality
discussed in the previous section also is an equivalence between
string vacua but it already holds in perturbation theory.
However, it also identifies different regions in the moduli 
space of string theories but these identifications do
not involve the string coupling or equivalently the dilaton.}
The concept of S--duality was first 
developed in four-dimensional
$N=4$ supersymmetric Yang-Mills 
theories which
is a somewhat special case 
since in terms of the above terminology 
one has $B=A$ or in other words a 
selfduality.\cite{mo-ol}$\!\!^{-}$\cite{mo-ol-su3,sen,schwarz-sen}
We start the discussion of S--duality with the 
$N=4$ Yang--Mills
theory\cite{harveyrev} 
and then briefly turn to the S--duality 
between the heterotic $SO(32)$ and the type I string.

\subsection{$N=4$ Yang-Mills theories in four space-time dimensions}

Extended supersymmetries are generated
by $N$ chiral charges $Q_\alpha^I$ and $N$
anti-chiral charges $Q_{\dot{\beta}}^I$ 
($I=1,\ldots ,N$)
which transform as Weyl spinors under the 
Lorentz group and  satisfy the 
algebra\footnote{We use the conventions 
$\alpha,\dot{\alpha}=1,2$ and thus the  
total number of supercharges is $4N$.\cite{we-ba}}
\be\label{eq:central}
\left\{ Q_\alpha^I, Q_{\dot{\beta}}^J\right\} 
= 2 \sigma_{\alpha\dot{\beta}}^\mu 
P_\mu\,  \delta^{IJ}\ ,\qquad
\left\{ Q_\alpha^I, Q_\beta^J\right\} = \epsilon_{\alpha\beta}\,  Z^{IJ}.
\ee

For $N=4$ 
the supersymmetric gauge multiplet contains a
vector boson $A_\mu$, four Weyl fermions $\chi_\alpha^{I}$
and six real scalar fields $\phi^{[IJ]}$ 
all in the adjoint 
representation of a gauge group $G$.
The scalar fields in the Cartan subalgebra of $G$ are flat directions
of the potential and thus at a generic point in their
field space 
$G$ is broken to its maximal Abelian subgroup 
$[U(1)]^{{\rm rank}(G)}$.
It has been shown that the field equations of
such theories have solitonic solutions which can be interpreted
as magnetic monopoles or more generally dyons -- 
states which carry both 
electric charge $q_e$ and 
magnetic charge $q_m$.\cite{thooft}\cto\cite{JZ}
Any two such states with charges $(q_e^1,q_m^1)$ and $(q_e^2,q_m^2)$
satisfy the Dirac-Zwanziger quantization 
condition \cite{dirac1}\cto\cite{dirac4}
\begin{equation}
q_e^1q_m^2-q_e^2q_m^1 = 2\pi k,
\end{equation}
with $k$ being integer.\footnote{
This condition ensures that one test
dyon with charges $(q_e^1,q_m^1)$ 
moving in the field of another dyon with charges
$(q_e^2,q_m^2)$ does not feel an Aharanov-Bohm 
effect.}
This condition is solved in terms of an elementary 
electric charge $e$ by \cite{mo-ol-su3}
\begin{equation}
q_e = n_e e - n_m e\, \frac{\theta}{2\pi} \ , \qquad q_m =  n_m\frac{4\pi}{e},
\end{equation} 
where $n_m$, $n_e$ are integers and $\theta$ is the 
coupling of the  $F\tilde F$ term (the $\theta$-angle).
Furthermore, the mass of any dyonic state satisfies
 the BPS bound \cite{bps1, bps2}
\begin{equation} \label{eq:bps}
M \geq \langle \phi \rangle  \sqrt{q_e^2 +q_m^2}\ .
\end{equation}

For a $U(1)$ gauge theory
the Maxwell equations in the vacuum
are invariant under the exchange 
${\bf E} \to - {\bf B}, {\bf B} \to  {\bf E}$ (or equivalently
under an exchange of the Bianchi identities with the field equation).
In the presence of magnetic monopoles this symmetry 
might be extended to the full Maxwell equations 
including matter.
Indeed,
Montonen and Olive conjectured that there exists such symmetry 
if the electric and magnetic quantum numbers are interchanged
and in addition the gauge coupling $e$ is
inverted according to \cite{mo-ol}
\begin{equation}\label{eq:strongweak}
e\to \frac{4\pi}{e}.
\end{equation}

More generally this symmetry can be combined
with the periodicity
of the $\theta$-angle $\theta \to \theta + 2\pi$.
Together they  generate
the discrete group $SL(2,Z)$ acting on the 
complex coupling \cite{FILQ,sen}
\begin{equation}    \label{eq:complex-gauge}
S := \frac{4\pi}{e^2} + i \frac{\theta}{2\pi}
\end{equation}
according to 
\begin{equation} \label{eq:sdual}
S \rightarrow \frac{aS-ib}{icS+d}\ ,\quad  ad -bc = 1,
\end{equation}
where $a,d,b,c\in {\bf Z}$. 
In terms of this more general symmetry operation
the quantum numbers $(n_e, n_m)$ should be relabelled
according to \cite{sen}
\begin{equation}\label{eq:sdualem}
\left(
\begin{array}{c}
n_e \\
n_m \end{array}\right)
\to 
\left(
\begin{array}{cc} a & b \\ c & d \end{array}\right)
\left(
\begin{array}{c}
n_e \\
n_m \end{array}\right)\ .
\end{equation}

One can also check that the BPS bound (\ref{eq:bps}) 
is invariant under the transformations
(\ref{eq:sdual}),(\ref{eq:sdualem}). 
This is of significance since the BPS-bound
is implied by the $N=4$ supersymmetry 
algebra 
and is believed to be  
an exact quantum formula (perturbatively and
non-perturbatively).\cite{mo-ol-su1}
To see this, one uses 
the $U(4)$ automorphism symmetry of the supersymmetry
algebra
to transform the matrix $Z^{IJ}$ of 
eq.~(\ref{eq:central}) into the form \cite{FSZ}
\begin{equation}
Z^{IJ} = \left(
\begin{array}{cc}
Z_1 \epsilon & 0 \\
 0 &Z_2\epsilon\end{array}\right) \ , 
\quad {\rm where}\quad \epsilon = \left(
\begin{array}{cc} 0 & 1 \\ -1 & 0 \end{array}\right).
\end{equation}
The positivity of the superalgebra  implies a
bound on the mass of all states
\begin{equation} \label{eq:sbps}
M\geq \left| Z_{1,2}\right| . 
\end{equation}
The massive representations of
the $N=4$ algebra  depend on
eq.~(\ref{eq:sbps}).\cite{FSZ}
First, there 
is the `long' massive multiplet with 256 states and 
maximal spin $s=2$.
The mass of this multiplet is strictly larger 
than any of the two central charges.
The `middle' multiplet contains 64 states of
maximal spin $s=3/2$ and its mass coincides with one of the
two central charges. Finally, the `short' multiplet
contains 16 states of
maximal spin $s=1$ with a mass that 
saturates  both central charges\footnote{
This is also the content of the massless
representation; for example the Yang-Mills
multiplet discussed at the beginning of this section
has a total of 16 states.}
\begin{equation}
M=\left| Z_1\right| =\left| Z_2\right|\ .
\end{equation}
The fact that these short multiplets
have fewer degrees of freedom 
implies a non-renormalization theorem for their 
mass and therefore the BPS states can be 
`followed' into the strong 
coupling regime.\footnote{Quantum corrections changing 
BPS states into non-BPS states
would imply that the number of degrees of freedom 
is changed by quantum corrections.}
This fact together with the $SL(2, Z)$
invariance of the BPS-bound (\ref{eq:sbps})
already supports the Montonen-Olive conjecture.
However, a further highly non-trivial test
of S--duality has been performed.\cite{sen} 
Starting 
from a massive, short BPS-multiplet $W^+$  
with charges $(n_e=1, n_m=0)$  S--duality 
predicts 16 dyonic multiplets with charges
\begin{equation}\label{eq:Sprediction}
\left(
\begin{array}{c}
n_e \\
n_m \end{array}\right)
= 
\left(
\begin{array}{cc} a & b \\ c & d \end{array}\right)
\left(
\begin{array}{c}
1 \\
0 \end{array}\right)
=
\left(
\begin{array}{c}
a \\
c \end{array}\right)\ 
\end{equation}
where
the constraint $ad-bc=1$ implies that $a$ and $c$ 
have to be
relatively primed. Indeed A.~Sen found the predicted
solitonic solutions for $(n_e={\rm odd}, n_m=2)$
with exactly the right multiplicity.\cite{sen}
All further tests so far confirm 
S--duality as an exact symmetry in $N=4$ Yang--Mills
theories.\cite{further0, further1, further2}

In order to discuss S--duality in string theory we have
to `embed' the previous discussion into $N=4$ supergravity.
The supergravity
multiplet contains one graviton $g_{\mu\nu}$, 
four gravitinos $\psi_{\mu\alpha}^I$, 
six Abelian vector bosons $\gamma_\mu^{[IJ]}$
called graviphotons, 
four Weyl fermions $\chi_\alpha^I$,
an antisymmetric tensor $b_{\mu\nu}$
and a real scalar $\D$.
In four space-time dimensions an antisymmetric tensor
contains only one physical degree of freedom
and  is dual to a real scalar
denoted by $a$.
It can be combined with the dilaton into the
complex scalar 
\begin{equation}
\frac{S}{4\pi} =  e^{-2\D} + i a\ .
\end{equation}
When coupled to a Yang-Mills vector multiplet
the coupling $S$ defined in eq.~(\ref{eq:complex-gauge})
is nothing but the VEV of this scalar $S$.
The classical equations of motion 
are invariant under an
$SL(2,R)$ acting on $S$ as in 
eq.~(\ref{eq:sdual}) with real coefficients but
space-time instantons break this continuous symmetry to 
the  discrete $SL(2,Z)$ symmetry.\cite{FILQ,rey,schwarz-sen}

$N=4$ supergravity coupled to 
Yang-Mills gauge multiplets arises in the low energy
limit of toroidally compactified  heterotic string theories. 
The bosonic part of the massless spectrum is obtained
by dimensional reduction of the ten-dimensional spectrum
displayed in table~1. One finds 
the graviton $g_{\mu\nu}$, 
the vector bosons $g_{\mu i}$, $b_{\mu i}$, $A_\mu^a$ and
the scalars $b_{\mu\nu} \sim a$, $\D$, $g_{ij}$,
$b_{ij}$, $A^a_i$
(as before Latin indices label compactified directions 
so for the case at hand we have $i,j = 1, \ldots, 6$).
The $A_\mu^a$ and $A^a_i$ combine into a non-Abelian
vector multiplet of either $E_8\times E_8$ or $SO(32)$
while the remaining degrees of freedom form the 
gravitational multiplet and six Abelian vector multiplets.
As we discussed in section~3, the scalars $A^a_i$
in the Cartan subalgebra are flat directions 
whose (generic)
VEVs break the gauge group to $[U(1)]^{16}$.
Thus at a generic point in field space there are 
 22 vector multiplets and one gravitational multiplet. 
The moduli space of the scalar fields is found
to be \cite{narain,schwarz-sen} 
\begin{equation}
{\cal M} = \left. \frac{SO(6,22)}{SO(6)\times SO(22)}\right/ \Gamma_T \times 
\left. \frac{SL(2,R)}{U(1)}\right/ \Gamma_S
\end{equation}
with the duality groups\footnote{Of course $\Gamma_T$ is consistent with the general 
formula (\ref{eq:Thettoroidal}).}
\begin{equation} \label{eq:S-group}
\Gamma_T = SO(6,22,Z)\ , \qquad
\Gamma_S = SL(2,Z) \ .
\end{equation}
$\Gamma_T$ only acts on the scalar 
moduli of the vector multiplets while $\Gamma_S$
transforms the two 
moduli of the gravity multiplet. 
In general there is a yet larger duality group
mixing all of the moduli termed U--duality;\cite{hull}
this will be the topic of section~5.

%
\subsection{S--duality between Type I and Heterotic $SO(32)$}
In string theory a
strong-weak coupling duality has been established between 
the type I theory  and the heterotic 
$SO(32)$ theory.\cite{various, type1a, type1b, polch-witt} 
First of all, from table~1 we immediately infer that
the massless spectra of heterotic $SO(32)$ 
and type I strings are identical.
However, the low energy effective actions 
are not the same in perturbation theory.
For the heterotic string one has
\begin{equation} \label{eq:het}
S_{heterotic} \sim \int d^{10}x \, \sqrt{-g}e^{-2\D }\,
\left\{ R + 4 \left(\partial \D \right)^2 - \frac{1}{4}Tr F^2 
- \frac{1}{12}H^2
\right\},
\end{equation}
where $H$ is the field strength for the antisymmetric tensor field $H=db$.
Type I string theory includes unoriented open and closed strings 
with $SO(32)$ Chan-Paton factors attached to the ends of the open string. 
The low energy effective 
action is made out of three contributions: 
the closed string NS-NS sector with the metric 
and the dilaton, 
the open string contribution resulting in 
the gauge kinetic term for the $SO(32)$ Yang-Mills fields, 
and the kinetic term for the 
antisymmetric tensor in the R-R sector
\begin{equation} \label{eq:typeI}
S_{type\, I} \sim \int d^{10}x \, \sqrt{-g} \left\{ e^{-2\D}
\left[ R + 4 \left( \partial \D \right)^2\right] -\frac{1}{4} e^{-\D} TrF^2
-\frac{1}{12} H^2\right\} .
\end{equation}
The different dilaton couplings arise because the second 
contribution comes from a disc diagram 
and the last contribution is a R-R term.
Replacing 
\begin{equation} \label{eq:het-1}
g_{\mu\nu} \rightarrow e^\D g_{\mu\nu} \ , \qquad \D \rightarrow -\D
\end{equation}
in the heterotic action (\ref{eq:het}) 
results in the type I action (\ref{eq:typeI}).
This suggests the following relation between the 
couplings 
\be
g_{\rm het} = e^{\D_{\rm het}} \sim 
g_{\rm I}^{-1} = e^{-\D_{\rm I}}\ .
\ee
In particular the strong coupling limit of the heterotic theory
is mapped to the weak coupling limit of the type I theory
and vice versa.

Further evidence of
this duality has been assembled.
In the type I theory there are one-branes and 
five-branes \cite{gimon} which can be mapped to 
the one- and five-branes
\cite{stromsoli,dufflu1,callan,duffPR} 
of the heterotic string.
The effective D-brane action is given by \cite{leigh}
\begin{equation}
S_{{\rm I}, p} \sim  \frac{1}{g_{\rm I}}\int d^{p+1}\eta \sqrt{-\det
g_{{\rm I}, p+1}}\ ,
\end{equation}   
where for simplicity we focus on vacua with constant dilaton and $H=F=0$. 
The coordinates
$\eta$ parameterize the $p$-brane world volume and 
$\det g_{{\rm I}, p+1}$ is
the
determinant of the induced metric.
Performing the duality transformation to the 
heterotic side (i.e.\ the
inverse of (\ref{eq:het-1})) gives
\begin{equation}
S_{{\rm het}, p} \sim g_{\rm het}^{\frac{1-p}{2}}
\int d^{p+1}\eta \sqrt{-\det g_{{\rm het}, p+1}}\ .
\end{equation}    
Thus, the one-brane of type I theory is 
mapped to a string
with tension of order $g_{\rm het}^0$ 
on the heterotic side.  This suggests
that the D-one-brane of type I
can be identified with the heterotic 
string.\cite{polch-witt}
The D-five-brane on the type I side corresponds to a five-brane with tension
of order $g_{\rm het}^{-2}$ on the heterotic side
which is  the  solitonic heterotic 
five-brane. 

Finally,  for compactifications  (including Wilson lines)
of both string theories to nine dimensions 
it has been observed that
points of enhanced gauge symmetry 
on the heterotic side correspond to
points where the perturbative description of 
type I theory breaks down.\cite{polch-witt}

%
%
\section{U--duality}
The concept of U--duality \cite{hull} 
was first discussed in 
$N=8$ supergravity in $d=4$ which has a 
total of 32 unbroken supercharges.
This theory only has a gravitational multiplet 
containing the  graviton $g_{\mu\nu}$,
eight gravitinos $\psi_{\mu\alpha}$ , 
28 graviphotons $\gamma_{\mu}$, 
56 spin-$\frac{1}{2}$ Weyl fermions
and 70 real scalar moduli $\phi$.
The 70 scalars parameterize the 
coset space \cite{cremmer}
\begin{equation}
{\cal M}_{\phi} = \frac{E_{7,7}}{SU(8)}\ ,
\end{equation}
where $E_{7,7}$ is a specific non-compact version of $E_7$.
The equations of motion are invariant under
$E_{7,7}({\bf R})$ and the electric and 
magnetic charges of the
28 graviphotons combine into 28 complex 
central charges of the $N=8$ superalgebra. 

$N=8$ supergravity arises as the low energy limit
of type IIA (or equivalently type IIB) compactified
on a six torus $T^6$. 
As before the bosonic part of the
massless spectrum can be
obtained by dimensional reduction of the ten-dimensional
spectrum listed in table~1.
From the NS-NS sector of type IIA 
one obtains  a graviton $g_{\mu\nu}$, 12 
graviphotons $g_{\mu i}$, $b_{\mu i}$ and 38 scalars 
$\D$, $b_{\mu\nu}\sim a$, $g_{ij}$, $b_{ij}$. 
In the R-R sector one finds  16 graviphotons $A_\mu$, 
$C_{\mu ij}$
and 32 scalars $A_i$ $C_{ijk}$, $C_{\mu\nu i}$.
The $g_{ij}$ and $b_{ij}$ are  the moduli of the 
six torus which parameterize the moduli space given
in eqs.~(\ref{eq:Mtoroidal}),(\ref{eq:Ttoroidal}).
Similarly, the dilaton-axion system spans a 
$SL(2,R)/U(1)$ coset divided by
the S--duality group (\ref{eq:S-group}).
So, altogether the NS-NS scalars live on 
the moduli space
\begin{equation}
{\cal M}_{\rm NS-NS} 
= \left. \frac{S(6,6)}{SO(6)\times SO(6)}\right/\Gamma_T \,
\times \, \left. \frac{SL(2,R)}{U(1)}\right/ \Gamma_S .
\end{equation}
It has been conjectured \cite{hull} that there is a much larger
duality group called U--duality which 
contains $\Gamma_S\times\Gamma_T$ 
as its maximal non-compact subgroup but transforms all scalars 
-- including the R-R scalars -- into each other.\footnote{This implies 
that the R-R scalars are charged under the U--duality group.}
For $N=8$ supergravity this U--duality group is conjectured to be 
\begin{equation}
\Gamma_U = E_{7,7}(Z)   
\end{equation}
so that globally the  moduli space is
\begin{equation}
{\cal M} = \left. \frac{E_{7,7}}{SU(8)}\right/ \Gamma_U\, .
\end{equation}

The conjectured presence of
this much larger quantum symmetry $\Gamma_U$
has been supported by a number of facts.\cite{hull,various}
First of all the BPS-bound is found to be 
invariant under $\Gamma_U$. 
Furthermore, the twelve graviphotons from the NS-NS
sector couple to `electrically' charged states of the
perturbative string spectrum. However, there can be no
R-R charges in the perturbative spectrum and thus
U-duality predicts the existence of non-perturbative
states which carry R-R charge.
Indeed,
the solitonic solutions of the field
equations come in representations of $\Gamma_U$
and R-R charged states 
have been constructed  as solitons 
and also as D-branes. 

A similar analysis has been performed for all toroidal compactifications
of type II string vacua and the result is listed in 
table~{\ref{tab:u}}.\cite{hull}
\begin{table}[t]
\caption{U--duality groups of toroidally compactified type II string theories.
}
\label{tab:u}
\begin{center}
\begin{tabular}{|l|l|l|l|}\hline
 $d$ & $G_{SUGRA}$ & $\Gamma_T$ & $\Gamma_U$ \\ 
\hline\hline
10 & $SL(2,R)$ & 1 & $SL(2,Z)$ \\
\hline
9 & $SL(2,R)\times SO(1,1)$ & $Z_2$ & $SL(2,Z)\times Z_2$ \\
\hline
8 & $SL(3,R)\times SL(2,R)$ & $SO(2,2,Z)$ & $SL(3,Z)\times SL(2,Z)$ \\
\hline
7 & $SL(5,R)$ & $SO(3,3,Z)$ & $SL(5,Z)$ \\
\hline
6 & $SO(5,5)$ & $SO(4,4,Z)$ & $SO(5,5,Z)$ \\
\hline
5 & $E_{6,6}(R)$ & $SO(5,5,Z)$ & $E_{6,6}(Z)$  \\
\hline
4 & $E_{7,7}(R)$ & $SO(6,6,Z)$ & $E_{7,7}(Z)$ \\ \hline
\end{tabular}\end{center}\end{table}
$G_{SUGRA}$ denotes the non-compact symmetry group
of the field equations in the 
corresponding supergravity theories.\cite{GZ}
Using U-duality Witten was able to give a complete picture of 
all possible strong coupling limits of type II vacua 
which further supports the validity of the conjecture.\cite{various}

For $d=10$ the situation is somewhat special
since type IIA and type IIB are perturbatively distinct
theories. The $d=10$ entry in table~3 refers to the type IIB string
which has a $SL(2,Z)$ acting on its two scalars 
$\D,\D'$.\cite{JSchwarz}
The ten-dimensional type IIA theory is discussed in the next section.

%
\section{M-theory} \label{sec:M-theory}
The discussion of the T-S-U-dualities and the resulting interrelation
between different string vacua led to the conjecture that
there is one underlying theory -- termed M-theory -- with
the perturbatively distinct string theories being just the various 
weak coupling limits in the moduli space of this fundamental M-theory.
In fact one more region of this moduli space has been
identified namely 11-dimensional supergravity.\cite{various}
Supergravity in 11 space-time dimensions is somewhat special 
since it is the highest possible 
dimension if one requires that there be no massless state
with spin higher than two.\cite{nahm}
The  field content of this theory contains the graviton $g_{\mu\nu}$, 
one gravitino $\psi_{\mu}$
and an antisymmetric  three-form $C_{\mu\nu\rho}$.\cite{cjs}

\subsection{Type IIA and  M-theory}
Type IIA supergravity can be constructed by simple dimensional reduction
of 11 dimensional supergravity.\cite{huq1}\cto\cite{huq3} 
The Kaluza--Klein 
modes $g_{\mu 11}$ and
the three-form $C_{\mu\nu\rho}$ give the R-R vector
$A_\mu$ and the R-R three-form  $C_{\mu\nu\rho}$, respectively.
The antisymmetric tensor $b_{\mu\nu}$ is obtained from $C_{\mu\nu 11}$
whereas the dilaton is related to the compactification radius specified
by $g_{11\, 11}$. Comparing the low energy effective actions
of the two theories similar to the  analysis of section~4.2
shows that the ten-dimensional type IIA string coupling $g_{A}$ and the
compactification radius $R$ are related by \cite{various}
\begin{equation}\label{eq:Rgrel}
R = g_{A}^{\frac{3}{2}}.
\end{equation}
Furthermore, 
the 11th component of the momentum operator $P_{11}$
appears as the central charge of the type IIA superalgebra
and  the Kaluza--Klein spectrum of the compactified 11 dimensional theory
corresponds to a tower of BPS-saturated states with masses
\begin{equation}
M_n \sim \frac{n}{g_{A}}\ ,
\end{equation}
where $n$ refers to the $n$th Fourier mode in the Kaluza--Klein expansion. 
In string perturbation theory
$g_{A}$ is taken to be small 
and hence the BPS states are very heavy and decouple. 
However, when the string coupling gets large 
the compactification radius $R$
also is large and the BPS states become light.  
Hence, in the strong coupling limit type IIA 
theory effectively decompactifies and the light
Kaluza--Klein spectrum of 11-dimensional supergravity
is nothing but a tower of type IIA BPS-states.
Another way  to put this is the statement that M-theory
compactified on a circle $S^1$ is type IIA string theory 
where the radius of this circle coincides with
the type IIA string coupling constants 
according to eq.~(\ref{eq:Rgrel}). 

%
\subsection{Heterotic $E_8\times E_8$ string from M-theory}
The  strong coupling limit of the
heterotic $E_8\times E_8$ string theory
is also related to 11-dimensional supergravity
but  this time not compactified on circle but rather
on a $Z_2$ orbifold of the circle, 
i.e.~$S_1/Z_2$.\cite{horava1}
The space coordinate 
$x^{11}$ is odd under the action of $Z_2$
and hence the three-form $C_{\mu\nu\rho}$
as well as $g_{\mu 11}$
are also odd. The $Z_2$ invariant spectrum 
in $d=10$ consists of the metric $g_{\mu \nu}$,
the antisymmetric tensor $C_{\mu\nu11}$
and the scalar $g_{11\, 11}$. 
Up to the gauge degrees of freedom
this is precisely the massless spectrum of the
ten-dimensional heterotic string. 
The $E_8\times E_8$ 
Yang-Mills fields arise in the twisted sector 
of the orbifold
compactification which so far cannot be
computed directly in M-theory but is deduced
in an indirect manner. 
The $Z_2$ truncation of 11 dimensional supergravity 
is inconsistent
in that it gives rise to gravitational anomalies.\cite{gaume} 
In order to cancel such 
anomalies non-Abelian gauge fields have to be present
in order to employ 
a Green-Schwarz mechanism.\cite{green-schwarz} 
Such additional states can only appear in the twisted sectors
of the orbifold theory which 
are located at the orbifold fixed points 
$x^{11}=0, x^{11}=\pi$. 
However, due to the $Z_2$ symmetry these two ten-dimensional hyperplanes
have to contribute equally to the anomaly.  
This can only be achieved for a gauge group which is a product
of two factors and thus $E_8\times E_8$ with one 
$E_8$ factor on each hyperplane is the only consistent candidate
for such a theory.\cite{horava1}

Exactly as in the type IIA case 
one has $R = g_{\rm het}^{\frac{3}{2}}$ and thus 
a small compactification radius gives the weakly
coupled heterotic string whereas the strongly coupled 
heterotic string
is equivalent to decompactified 11-dimensional supergravity.
Using the previous terminology the heterotic $E_8\times E_8$
string theory can be viewed as M-theory compactified 
on $S_1/Z_2$.

Let us summarize the situation so far.
In ten dimensions the heterotic $SO(32)$ and the type I
string theories are
S-dual,  that is they are quantum equivalent
and merely two different perturbative expansions
of the same quantum theory. 
The type IIB theory is selfdual with a strong coupling
limit governed by $SL(2, Z)$.
The strong coupling limits of type IIA and 
the heterotic $E_8 \times E_8$ theory are 11-dimensional
supergravity and they can be viewed as circle or orbifold
compactifications of M-theory.

In section~3 we recalled that below $d=10$ 
the type IIA and type IIB theory are in
the same moduli space and similarly 
the heterotic $E_8 \times E_8$
and $SO(32)$ theories are in the same moduli space.
Thus, 
if one also treats the radii of the 
toroidal compactifications
as parameters
there now is an intriguing
relation between all five perturbative string theories
and it has been conjectured that all of them
are just different weak coupling limits in the moduli space
of one underlying (fundamental) theory called 
M-theory.\cite{townsendrev}

%
\section{String compactifications with broken
supercharges in six space-time dimensions}

All string vacua considered so far have either 32 or 16 
supercharges and they are toroidal compactifications of
ten-dimensional $N=2$ or $N=1$ supergravities
which leave all supercharges intact.
In lower space-time dimensions it is possible to consider
compactifications which break some of the supercharges.
Such compact manifolds which at the same time preserve the
consistency of the field equations of string theory
are known as Calabi--Yau manifolds.\cite{tex1}

%
\subsection{The $K3$ surface}
A Calabi-Yau  manifold $Y$ is a Ricci-flat K\"ahler manifold
of vanishing first Chern class with 
holonomy group $SU(n)$ where $n$ is
the complex dimension of $Y$. 
A (complex) one dimensional Calabi-Yau manifold is  
topologically always
a torus. In two complex dimensions  
all Calabi-Yau manifolds are topologically
equivalent to  the $K3$ surface.\cite{aspinwallrev}
The moduli 
space of non-trivial metric deformations on a $K3$ 
is 58-dimensional and given by the coset space 
\begin{equation}
{\cal M}\ =\  R^+\ \times\
\left.\frac{SO(3,19)}{SO(3)\times SO(19)}\right/ SO(3,19,Z)\ ,
\end{equation}
where the second factor is the Teichm\"uller space 
for Einstein metrics of
volume one on a $K3$ surface and 
the first factor is associated with the
size of the $K3$.\cite{aspinwallrev}

In addition to the metric deformations there are
non-trivial two-forms on  $K3$ which are related
to the antisymmetric tensor of string theory.
On any K\"ahler manifold, the differential k-forms 
can be decomposed into $(p,q)$-forms with $p$ holomorphic
and $q$ antiholomorphic differentials.
The harmonic $(p,q)$-forms form the cohomology groups 
$H^{p,q}$ of dimension $h^{p,q}$ and for  $K3$ one has
\begin{equation} \label{eq:hodge}
\begin{array}{c c c c c}
            &              &  h^{0,0} &            &            \\
            & h^{1,0}\! &              &\! h^{0,1}\!&            \\
  h^{2,0}\! & \!             & h^{1,1} & \!            &\! h^{0,2}\\
            & h^{2,1} &              &h^{1,2}&             \\
             &              &h^{2,2}&              & 
\end{array} =
\begin{array}{c c c c c}
  &   & 1   &  & \\
  & 0\! &      &\! 0& \\
1\! &\!   & 20 & \! &\! 1\\
  & 0\! &      &\! 0&  \\
  &   &1    &  & 
\end{array}\  .
\end{equation}
Thus there are 22 harmonic $p+q=2$-forms which represent the 
non-trivial deformations of the antisymmetric tensor 
$b_{ij}$.
Together with the 58 deformations of the 
metric they form
the 80-dimensional moduli space 
of $K3$ string compactifications \cite{NS,aspinwall}
\begin{equation}
{\cal M} = 
\left. \frac{SO(4,20)}{SO(4)\times SO(20)}\right/ 
SO(4,20,Z)\ .
\end{equation}

For later reference let us also record
that the Euler number of $K3$ is found to be 
\begin{equation}
\chi = \sum_{p,q} (-)^{p+q} h^{p,q} = 24.
\end{equation}
%
%
\subsection{Supergravities on $K3$}
Whenever the space-time dimension obeys
$d=4k+2$ ($k$ being an integer) the left and right handed
spinors are independent and as a consequence one can
have chiral supersymmetries.
In $d=10$ the irreducible spinor representations
are Majorana-Weyl and the $N=2$ supersymmetry
either has two gravitinos
of the same chirality (type IIB) or of opposite 
chirality (type IIA).
In $d=6$ the irreducible spinor representations
are symplectic Majorana (or pseudo-real) 
and again one can have chiral and non-chiral supersymmetries.
The toroidally compactified type II string
corresponds in $d=6$ to a non-chiral supergravity
with 16 chiral and 16 anti-chiral supercharges.
Such theories have two chiral and two anti-chiral
gravitinos and are often denoted as $(2,2)$ supergravity.

Compactification on $K3$ breaks half of the supercharges
and one finds the  two possibilities of having
either 16 chiral supercharges (two chiral gravitinos)
or 8 chiral plus 8 anti-chiral supercharges 
(one chiral and one anti-chiral gravitino).
The first possibility arises when type IIB is compactified
on $K3$ and thus this supergravity is also called 
type IIB or $(2,0)$. 
The second case occurs as a compactification of type IIA
on $K3$ is denoted type IIA or $(1,1)$ supergravity.
Finally, compactification of the heterotic or type I
string theory on $K3$ only leaves 8 supercharges
(one gravitino) and is called $(1,0)$ supergravity.

%
\subsection{Type IIB compactified on K3}
The chiral $(2,0)$ supergravity has a
gravitational multiplet containing as boso\-nic components
the graviton $g_{\mu\nu}$ and  five antisymmetric
tensor fields $b_{\mu\nu}^{+}$ whose field strength 
is selfdual.
The only other possible multiplet in this supergravity
is a tensor multiplet
containing an antisymmetric tensor
$b_{\mu\nu}^-$ with an anti-selfdual field strength and
five scalars $\phi$.
Compactifying the type IIB string on $K3$ leads to
the massless modes 
$g_{\mu\nu}, b_{\mu\nu}, \D, g_{ij}, b_{ij}$
from the NS-NS sector and 
$b_{\mu\nu}', \D^\prime, b_{ij}^\prime, 
C_{\mu\nu\rho\sigma},C_{\mu\nu ij}$
from the R-R sector where 
$g_{ij}$ denote the 58 zero modes of the metric on $K3$
and the $b_{ij}, b_{ij}'$ each are  22 harmonic two-forms.
In $d=6$ a four form has only one physical
degree of freedom and is dual to a real scalar
$C_{\mu\nu\rho\sigma}\sim a$. Furthermore, the 22 antisymmetric
tensor fields $C_{\mu\nu ij}$ can be decomposed into
three selfdual and 19 anti-selfdual tensors corresponding
to an analogous decomposition of the 
22 two-forms on $K3$.\cite{aspinwallrev}
So altogether there are 81 NS-NS and 24 R-R scalars,
5 selfdual and 21 anti-selfdual tensors which 
altogether combine into one gravitational 
and 21 tensor multiplets.
This is precisely the combination of an anomaly free
type IIB spectrum.\cite{townsend}
The scalars parameterize the 
105-dimensional moduli space \cite{romans,aspe,hul}
\begin{equation}
{\cal M} = \left. \frac{SO(5,21)}{SO(5)\times SO(21)}\right/ \Gamma_U,
\end{equation}
with a  U--duality group $\Gamma_U = SL(5,21,Z)$.
One of the strong coupling limits of this theory is 
again 11-dimensional supergravity,\cite{dasgupta,wittenm}
 this time 
compactified on $T^5/Z_2$.\footnote{
Compactifying 11-dimensional supergravity
on an orbifold $T^5/Z_2$ gives chiral $(2,0)$ 
supergravity with one gravity multiplet and 
five tensor multiplets  from the untwisted sector. 
The twisted sector is again inferred by 
anomaly cancellation and provides 16 further 
tensor multiplets. The weakly coupled
type IIB theory on $K3$ corresponds to a `smashed' 
$T^5/Z_2$ where the
32 fixed points degenerate into 16 pairs 
and the 16 tensor multiplets are
equally distributed among those pairs.}

%
\subsection{Type IIA compactified on K3}
The non-chiral type IIA $(1,1)$ supergravity has
a gravitational multiplet as well as vector multiplets. 
The bosonic fields in the
gravity multiplet are the graviton $g_{\mu\nu}$, an antisymmetric 
tensor $b_{\mu\nu}$, 
four  graviphotons  $\gamma_\mu$
 and one real scalar $\D$.
The vector multiplets contain a vector boson $A_\mu$ 
and four scalars. 

Compactification of the type IIA string on $K3$
results in the following bosonic spectrum: 
the graviton $g_{\mu\nu}$, an antisymmetric tensor 
$b_{\mu\nu}$, the dilaton $\D$, 
58 scalars $g_{ij}$ and  22 scalars $b_{ij}$ all 
from the NS-NS sector. In the R-R sector there is
one vector $A_\mu$, an additional  vector arising
as the dual of the three-form 
$C_{\mu\nu\rho}\sim A_\mu^\prime$ and 
22 vectors $C_{\mu ij}$. These fields combine
into one gravity multiplet and 20 vector multiplets.
The 81 scalars, (all of them come from the NS-NS 
sector since $h_{p,q}=0$ for
$p+q$ odd), parameterize the moduli space \cite{NS,aspinwall}
\begin{equation}  \label{eq:2aonk3}
{\cal M}\ =\ R^+\ \times \ 
 \left. \frac{SO(4,20)}{SO(4)\times SO(20)}\right/ \Gamma_T
\end{equation}
where the T--duality group is $\Gamma_T = SO(4,20,Z)$.

%
\subsection{Heterotic strings on $T^4$} 
Before we discuss $K3$ compactifications 
of the heterotic string we need to pause and first 
reconsider  
toroidal compactification of the heterotic string.
Such string vacua also have 16 supercharges and 
the low energy limit  is the non-chiral type IIA or
$(1,1)$ supergravity.
The bosonic spectrum is again obtained by dimensional
reduction from the ten-dimensional heterotic spectrum
of table~1.
At a generic point in the moduli space (where the non-Abelian gauge symmetry is broken to $[U(1)]^{16}$)
one has the graviton $g_{\mu\nu}$, an antisymmetric tensor 
$b_{\mu\nu}$, the dilaton $\D$, 80 scalars 
$g_{ij}, b_{ij}, A_i^a$ and 24 vectors 
$A_\mu^a,g_{\mu j}, b_{\mu j}$.
Together they form one gravitational multiplet and 
20 vector multiplets exactly as for type IIA
compactified on $K3$.
Furthermore, from eqs.~(\ref{eq:Mhettoroidal}), 
(\ref{eq:2aonk3})
we learn that also the moduli spaces of the two
string compactifications coincide.\footnote{
In addition, the existence of a solitonic 
string suggested that there is a description
in terms of elementary 
five-branes.\cite{stromsoli}\cto\cite{ruiz2}
 In six dimensions
this implies a string/string duality.\cite{DK}}
This led to the 
conjecture that type IIA on $K3$
and the heterotic string on $T^4$ are possibly
quantum equivalent  or
S-dual to each other.\cite{DK}\cto\cite{hull,duff,various} 

The effective actions of the two perturbative theories
can be compared using the methods of section~4.2.
They turn out to agree if one  identifies
\cite{various}
\bea \label{eq:strowe}
\D_{\rm het} &=& - \D_{A}\ ,\nonumber \\
H_{\rm het} &=& e^{-2\D_{A}} *H_{A}\ ,  \\
(g_{\rm het})_{\mu\nu} &=& e^{-2\D_{A}} (g_{A})_{\mu\nu}\ ,
\nonumber 
\eea
where $H=db$ is the field strength of the antisymmetric
tensor and $*H$ is its Poincare dual.
The first equation in (\ref{eq:strowe})
again implies a strong-weak coupling relation while
the second  is the equivalent of an 
electric-magnetic duality.
Further evidence for this S-duality 
arises from the observation that the zero modes 
in a solitonic string background
of the type IIA theory compactified on $K3$ 
have the same structure
as the Kaluza--Klein modes 
of the heterotic string compactified 
on $T^4$.\cite{sen-he, hav-stro}

However, there is an immediate puzzle. 
We know that on special subspaces of the moduli space of
the toroidally compactified heterotic strings 
the gauge symmetry can be enhanced to 
non-Abelian gauge groups. 
At face value it seems impossible 
to  obtain such
a gauge symmetry enhancement in the  type IIA theory
where all vectors are in the R-R sector with
no massless charged states possible. 
This question leads  to another important 
topic in string dualities --
the study of singularities or rather 
singular couplings in the moduli space.

%
\subsection{Singularities in the moduli space}
Singularities  in the moduli space of string vacua
can in principle be of two  different origins.
First, the singularities might
be an artifact of perturbation theory and 
smoothed out once quantum corrections are 
taken into account. 
The other possibility is that a singularity is physical
and signals the breakdown of some approximation.
The prime example for this latter case is 
the appearance of additional massless degrees
of freedom on a subspace of the moduli space.\footnote{Almost
everywhere on the moduli space these degrees of freedom 
are heavy and 
thus have been integrated out of the effective theory.
However, if their masses become small somewhere in the moduli space
the approximation of integrating out such degrees of freedom
is not valid on the entire moduli space.
The breakdown of this approximation manifests itself as a
singularity in some of the couplings of the effective
theory.}
A well known example occurs in  four-dimensional non-Abelian 
(supersymmetric) gauge theories with light and heavy degrees
of freedom. 
Below the threshold scale $M$ of the heavy states the gauge couplings of 
the light modes obey
\begin{equation} \label{eq:peterfunktion}
g^{-2}_{\rm low} = g^{-2}_{\rm high} + c \log M,
\end{equation}
where $c$ is some model dependent constant and $g_{\rm high}$ 
is the  coupling  above the heavy threshold. 
In string theory the mass often  is a function of the
moduli $M(\phi)$ with a zero somewhere in the moduli space
$M(\phi_0)=0$.
At such points the gauge coupling $g^{-2}_{\rm low}$ becomes
singular due to the inappropriate approximation 
of integrating out the heavy states.\footnote{Quantum 
mechanically the number and position of the 
singularities and 
the interpretation of the fields becoming massless at the
singularities can be quite different 
to the classical picture.\cite{seiwi1}}

Indeed the $K3$ surface has
orbifold singularities (following an A-D-E classification)
whenever  a two-cycle of the $K3$ shrinks to zero. 
A careful analysis  shows that in 
$K3$ compactifications of type IIA the origin
of these 
singularities can be interpreted as 
non-Abelian gauge bosons  
(of an A-D-E gauge group)
becoming massless.\cite{various,aspinADE}
Thus a point of non-Abelian gauge
enhancement on the heterotic side corresponds to an 
orbifold singularity 
on the type IIA side.\footnote{This phenomenon can also be 
viewed as the type IIA two-brane 
wrapping around the 2-cycles of the $K3$.\cite{stromingerBH}}

Although the original puzzle about the subspaces
of enhanced gauge symmetries in type IIA is resolved
the type IIB string compactified on $K3$ now raises
a  puzzle. Here the same singularities are present
but the $(2,0)$ supergravity does not have 
any vector multiplets and hence it is impossible to
explain the singularities by additional 
massless gauge bosons. Instead, 
it has been argued that in type IIB the singularities 
are caused by extended objects --
non-critical strings  -- becoming tensionless at the
orbifold singularities of  $K3$.\cite{wi-tension} 
Those tensionless strings
arise when a selfdual three brane of the ten-dimensional 
type IIB theory
wraps around a vanishing two cycle of the $K3$.
%
\subsection{The heterotic string compactified on K3}
Compactifying the heterotic string on  $K3$ only leaves 8 supercharges
intact which is the minimal or $(1,0)$ supergravity in $d=6$.
This theory has  four distinct supermultiplets.
The gravitational  multiplet contains 
the graviton $g_{\mu\nu}$ and an anti-selfdual two 
form $b^-_{\mu\nu}$ as bosonic components.
The tensor multiplet has a selfdual two-form $b_{\mu\nu}^+$ 
and a scalar $\D $,
the vector multiplet features  only a vector $A_\mu$
and finally the  hyper multiplet contains  four scalars 
$q$ as bosonic components.
This supergravity is chiral and thus gauge and gravitational
anomaly cancellation imposes constraints on the allowed spectrum.
The anomaly can be characterized by 
the anomaly eight-form $I_8$ which is exact 
($dI_8=0$) and invariant
under general coordinate and gauge 
transformations.
$I_8$ can be written as an exterior derivative of
a seven-form $I_8 = dI_7$
whose variation obeys $\delta I_7 = d I^\prime_6$ with 
$I^\prime_6$ being the anomaly ($\delta S = \int I^\prime_6$).
The generic form\footnote{We exclude
Abelian gauge factors for simplicity.}
 of $I_8$ is given by  \cite{gaume}
\begin{equation}
I_8\, =\, \alpha\, tr R^4 +\, \beta\, \left(tr R^2\right)^2 
+\, \gamma\, tr R^2 tr F^2 +\, \delta\, \left(tr F^2\right)^2 ,
\end{equation}
where $R$ is the curvature two-form,
$F$ is the Yang-Mills two-form
and $\alpha,\ldots,\delta$ are real coefficients
which depend on the spectrum of the theory. 
The anomaly can only be cancelled if 
$\alpha$ vanishes.
One finds \cite{gswest}\cto\cite{schwarzanom}
\begin{equation}\label{eq:Rfourconstraint}
\alpha = n_H - n_V + 29 n_T -273 \stackrel{!}{=} 0,
\end{equation}
where $n_H$, $n_V$ and $n_T$ are the numbers of hyper, vector and
tensor multiplets, respectively. In the perturbative spectrum of the 
heterotic string there is only one tensor multiplet 
(from $b_{\mu\nu}$) and hence anomaly cancellation demands
\begin{equation} \label{eq:hyper-vector}
n_H-n_V=244\ .
\end{equation}
The remaining anomaly eight form has to factorize in order to employ a
Green--Schwarz mechanism. More precisely, one needs
$I_8 \sim  X_4 \wedge \tilde{X}_4$
where
\be\label{eq:xfour}
X_4 = tr R^2 - \sum_a v_a\left( trF^2\right)_a\ ,\qquad
\tilde{X}_4 = tr R^2 - \sum_a \tilde{v}_a \left( tr F^2\right)_a\ .
\ee
The index $a$ labels the factors $G_a$ of the gauge group
$G= \otimes_a G_a$ and  $v_a, \tilde{v}_a$ are constants 
which depend on the
massless spectrum.\cite{gswest}\cto\cite{schwarzanom}
\setcounter{footnote}{0}
In the Green--Schwarz mechanism one defines a modified field strength $H$
for the antisymmetric tensor
\begin{equation} \label{eq:chern}
H = db +\omega^L - \sum_a v_a\, \omega_a^{YM},
\end{equation}
such that $dH = X_4$.
($\omega^L$ is a Lorentz--Chern--Simons term and $\omega_a^{YM}$ 
is the Yang--Mills Chern--Simons term.)
Adding to the action the Green-Schwarz counterterm
\begin{equation} \label{eq:green-schwarz}
\int_{R^6}b\wedge \tilde{X}_4
\end{equation} 
finally results in a complete anomaly cancellation.\cite{gswest}

In order to ensure a globally well-defined $H$ on the compact 
$K3$ the integral $\int_{K3} dH$ has to vanish. 
Using eqs.~(\ref{eq:xfour}), (\ref{eq:chern}) this implies
\begin{equation}\label{eq:instanton}
\sum_a n_a\, \equiv\, \sum_a \int_{K3}\left( trF^2\right)_a
\,=\, \int_{K3}trR^2\, =\, 24,
\end{equation}
where the last equation used the fact that 
24 is the Euler number of $K3$.
From eq.~(\ref{eq:instanton}) we learn that the heterotic string
compactified on $K3$ 
necessarily needs a non-vanishing instanton number
$n_a$.

As a first example let us consider the $E_8\times E_8$ theory with instantons
only in one of the $E_8$ factors breaking that $E_8$ completely and leaving the
other $E_8$ factor unbroken. The dimension of the moduli space of
instantons on $K3$ in a group $G$ with dual Coxeter number $h$ 
and instanton number $n$ is given by 
\begin{equation}
dim {\cal M}_n =4 (n h- dim (G)) \ .
\end{equation}
For the case at hand we have $dim(G) = dim(E_8)= 248,\,
h=30,\, n=24$ and thus 
$dim {\cal M}_{24} = 4\cdot 472$. 
As we discussed at the beginning of this section, 
hypermultiplets contain four real scalars each, 
and thus the instanton moduli space is parameterized by
472 hypermultiplets. The heterotic string vacua have an additional 20 
hypermultiplets which host the 80 moduli of $K3$. 
Thus, altogether one has 
\begin{equation}
n_H= 472 + 20 = 492\ , \qquad  n_V=dim(E_8)=248\ ,
\end{equation}
which indeed satisfies eq.~(\ref{eq:hyper-vector}).\footnote{ 
By aligning some of the 24 
instantons, i.e.\ going to special points of the moduli space,
the gauge group can be enhanced up to at most $E_8\times E_7$
when all 24 instantons sit in an $SU(2)$.}

As a second example we consider again the 
$E_8\times E_8$ heterotic string
but this time with $n_1$ instantons in one $E_8$ and 
$n_2$ instantons in
the other $E_8$ factor. Furthermore, for $n_a \geq 10$ 
the gauge group is  completely broken at a generic point in 
the moduli space and 
one has $dim {\cal M} = 4 (720-2\cdot248)= 4\cdot 224$.
Thus, $n_H =224 + 20 = 244, n_V=0$ in 
agreement with (\ref{eq:hyper-vector}).  
At special points in the moduli space 
an enhanced gauge group can open up 
which is  at most $E_7\times E_7$ when all
instantons sit in $SU(2)\times SU(2)$.

Finally, for $n_1=8, n_2=16$ the gauge group is $SO(8)$
at a generic point in the moduli space and can be enhanced
to $SO(28)\times SU(2)$  at a special point. 
It has been shown that this
$E_8\times E_8$ heterotic vacuum is quantum equivalent
to the compactification of the $SO(32)$ heterotic string
on $K3$.\cite{aldazabal,morva1,AGK3}
Thus, also for $K3$ compactifications the two 
heterotic string theories sit in one and the same
moduli space.

In all of these examples the kinetic energy of the gauge fields is strongly
constrained by supersymmetry.
The dilaton is a member of the tensor multiplet which also 
contains the (selfdual) part of the antisymmetric tensor.
Its couplings to the vector multiplets
are fixed by the Chern--Simons interactions (\ref{eq:chern})
which in turn also fixes the interactions of the dilaton 
with the gauge fields \cite{sagnotti}
\begin{equation} \label{eq:kinetic}
{\cal L} \sim \sqrt{g} \sum_a \left( v_a e^{-2\D} +\tilde{v}_a\right)
\left( tr F_{\mu\nu}F^{\mu\nu}\right)_a \ .
\end{equation}
For the second example discussed above one finds the 
coefficients \cite{erler,schwarzanom}
\begin{equation}
v_1 = v_2 = \frac{1}{6}\ , \qquad
\tilde{v}_1 = \frac{\left( n_1 -12\right)}{6} \ , \quad
\tilde{v}_2=\frac{\left( n_2 - 12\right)}{6} \ .
\end{equation}
For $n_1\neq n_2$ one of the $\tilde{v}$ is necessarily negative
and hence the gauge kinetic term (\ref{eq:kinetic}) changes sign at
\begin{equation} \label{eq:phase}
e^{-2\D} = -\frac{\tilde{v_1}}{v_1} = 12 - n_1,
\end{equation}
(where we have arbitrarily  chosen $n_1$ to be less then 12). 
Eq.~(\ref{eq:phase}) implies a singularity  at strong coupling. 
Similar to the case of the type IIB string compactified
on $K3$ it has been argued that this singularity
originates from a non-critical string with a tension 
controlled by the dilaton, becoming  
tensionless.\cite{nati,lue} 

Apart from this strong coupling singularity 
also the instanton moduli space typically develops singularities
when the size $\rho$ of an instanton   
approaches zero.\cite{callan,wi-tension}
These singularities necessarily are weak coupling singularities
since the dilaton resides in the tensor multiplet
which can have no gauge neutral couplings with the moduli 
in hypermultiplets and
thus at $\rho \to 0$ the dilaton can be arbitrarily weak.
It has been shown that the conformal field theory
description nevertheless breaks down and the 
space-time geometry develops a semi-infinite tube with the
dilaton becoming large `further down the tube'.\cite{callan}
For the $SO(32)$ heterotic string Witten gave convincing
evidence that at the locus of the collapsing instanton 
additional vector bosons become massless corresponding to
a non-perturbative enhancement of the gauge group.\cite{schmal}
For a single instanton the non-perturbative gauge group 
is $SU(2)$ while $k$ instantons 
collapsing at distinct points of the instanton moduli
space enlarge the perturbative gauge group 
to $G=SO(32)\times SU(2)^k$ with $k$ additional 
hypermultiplets in the $({\bf 32},{\bf 2})$ representation
of $G$.
$k$ instantons collapsing to the same point yield 
$G=SO(32)\times Sp(k)$ with one additional 
hypermultiplet in the $({\bf 32},{\bf 2k})$ 
representation
of $G$.
Thus, for 24 instantons the maximal gauge group possible
is $G = SO(32)\times Sp(24)$.
The enhancement factor $Sp(24)$ is completely invisible
in string perturbation theory and it violates
the perturbative bound on the rank of G
set by the central charge of the underlying CFT.

Further evidence of this picture arises from the 
heterotic -- type I S--duality which relates
a  small heterotic instanton to
a D-five-brane carrying $SU(2)$ Chan--Paton factors. 
Whenever $k$ of
such D-branes sit on top of each other the gauge symmetry is enhanced
to $Sp(k)$.\cite{schmal} 

For the $E_8\times E_8$ heterotic string a similar 
mechanism
is inconsistent. Instead it has been argued that
at the locus of the collapsing instanton
an entire non-critical string becomes 
tensionless.\cite{nati,hanany}
One of the massless modes of this string is an additional
tensor multiplet containing a real scalar as its lowest
component. This scalar can be viewed as parameterizing a new
(Higgs) branch in the moduli space. However, consistency 
requires eq.~(\ref{eq:Rfourconstraint}) to be satisfied
and thus on the new branch there necessarily is 
a different spectrum. For example for $n_T=2, n_V=0$
one has $n_H=244-29=215$.
This branch of moduli space is again invisible 
in heterotic
perturbation theory where only $n_T=1$ is possible.
However, the non-perturbative properties
of these heterotic vacua have been captured by 
constructing  appropriate weakly coupled dual 
vacua. These constructions go under the name of  
F-theory and therefore we pause for
a moment in order to have a brief look 
at F-theory.\cite{vafas-f,morva1}

\subsection{F-theory}
The type IIB theory in ten space-time dimensions
has an $SL(2,Z)$ quantum symmetry acting as in 
eq.~(\ref{eq:sdual}) on the complex 
\begin{equation} \label{eq:axion-dilaton}
\tau = e^{-2\D}  + i \D^\prime\ ,
\end{equation}
where $\D$ and  $\D^\prime$ are the two scalar fields of
type IIB theory (c.f.~table~1).
This fact led Vafa to propose that the type IIB string 
could be viewed as the toroidal compactification
of a twelve-dimensional theory, called F-theory,
 where $\tau$ is the 
complex structure modulus of the two-torus 
$T^2$ and the K\"ahler class
modulus is frozen.\cite{vafas-f}
Apart from having a geometrical interpretation of 
the $SL(2,Z)$ symmetry
this proposal led to the construction
of new, non-perturbative string vacua in 
lower space-time dimensions.
In order to preserve the 
$SL(2,Z)$ quantum symmetry the 
compactification manifold cannot be arbitrary
but has to be what is called an elliptic fibration.
That is, the manifold is locally a fibre bundle
with a two-torus $T^2$ over some base $B$ but 
with
a finite number of singular points.
As a consequence non-trivial closed loops 
on $B$ can induce a non-trivial $SL(2,Z)$
transformation of the fibre.
This implies that the dilaton is not constant 
on the compactification manifold but can `jump'
by an $SL(2,Z)$ transformation.\cite{stringy-cosmic}
It is precisely this fact which results in
non-trivial (non-perturbative) string vacua
inaccessible in string perturbation theory. 

For example F-theory compactified on a elliptic
$K3$ yields
an 8-dimensio\-nal vacuum with 16 supercharges
which is quantum equivalent to the heterotic string
compactified on $T^2$.\cite{vafas-f,sen-f}
On the other hand F-theory compactified on an elliptic
Calabi--Yau threefold has 8 unbroken supercharges and
is quantum equivalent to
the heterotic string compactified on $K3$.\cite{morva1}
In fact there is a beautiful correspondence
between the heterotic vacua labelled by the 
instanton numbers $(n_1,n_2)$ and 
elliptically fibred Calabi-Yau manifolds 
with the base being the
Hirzebruch surfaces $I\!\! F_{n_2-12}$
(we have again chosen $n_2\ge n_1$).\cite{morva1}
These F-theory vacua capture the non-perturbative physics
of the heterotic string including the possibility
of additional tensor multiplets, the transitions
between the various branches of moduli space
and subspaces of symmetry 
enhancement.\cite{morva1,AG,BIKMSV}

%
\subsection{Heterotic -- heterotic duality}
The string--string duality\cite{DK} in $d=6$ which 
(in part) is responsible
for the S-duality of the heterotic string on $T^4$
and the type IIA on $K3$ also has 
a somewhat surprising manifestation 
in $K3$ compactifications of the heterotic string.
For $n_1=n_2=12$ eq.~(\ref{eq:phase}) reveals
that there is no strong coupling singularity
so that this class of heterotic vacua is well
defined for all values of the dilaton.
Furthermore, it has been shown 
that there is a selfduality among the
$n_1=n_2=12$ heterotic vacua.\cite{duffmewi,morva1,AG}
 More precisely, 
the theory is invariant under 
\bea
\D &\to & - \D \ , \qquad
g_{\mu\nu}\to  e^{-\D}g_{\mu\nu} \nonumber\\
H & \to & e^{-\D} * H\ , \qquad
X_4 \leftrightarrow \tilde{X}_4
\eea
if in addition perturbative gauge fields
are replaced
by non-perturbative gauge fields
and the hypermultiplet moduli spaces are mapped
non-trivially onto each other.
Curiously, this duality requires the existence of 
non-perturbative gauge fields with exactly 
the properties discussed in section~7.7 for 
$SO(32)$ heterotic strings.
This posed a slight puzzle since
for $E_8\times E_8$ vacua the singularity
of small instantons is caused by a non-critical string
turning tensionless rather than additional
massless gauge bosons. However, by mapping the
$n_1=n_2=12$ heterotic vacuum 
to a particular type I vacuum \cite{gimon}
which indeed does have non-perturbative gauge
fields this issue has been resolved and 
the gauge fields are  shown to exist
also for this class of heterotic vacua.\cite{bele}

%
\section{String compactifications with broken supercharges
in four space-time dimensions}
In this last section we focus on string compactifications 
with 8 unbroken supercharges in $d=4$ or in other words
on string vacua with  $N=2$ supergravity.
Such vacua arise either from the type II string 
compactified on a Calabi--Yau threefold or
from the heterotic string compactified on 
$K3\times T^2$.\footnote{Such vacua and an 
expanded treatment of the content
of this section including a more complete list
of references can be found in a recent 
review.\cite{kristin}}
Also these two classes of 
string vacua are believed to be quantum
equivalent.\cite{kachru,fhsv}

\subsection{$N=2$ supergravity}
In $N=2$ the gravitational multiplet contains the graviton $g_{\mu\nu}$,
two gravitini $\psi^{I}_{\mu\alpha}$  and an Abelian
graviphoton $\gamma_\mu$. 
One also has  vector multiplets $V$ 
with a gauge field $A_\mu$, 
two gauginos $\lambda_{\alpha}^I$ and a complex scalar $\phi$
as well as hypermultiplets with two Weyl spinors $\chi^{I}_\alpha$ 
and four real scalars $q^{IJ}$. In addition there are 
three distinct multiplets containing an antisymmetric tensor;
the vector-tensor multiplet  $VT$ 
features an Abelian gauge field $A_\mu$,
an antisymmetric tensor $b_{\mu\nu}$, two Weyl fermions $\chi^I_\alpha $ 
and one real scalar $\D$.
The tensor multiplet $T$ has an antisymmetric tensor $b_{\mu\nu}$,
two Weyl spinors $\chi_{\alpha}^I$, a complex scalar $\phi$ and a real
scalar $\D$.
Finally the double tensor multiplet $\Pi$ contains two antisymmetric
tensors $b_{\mu\nu}$, $b^{\prime}_{\mu\nu}$, 
two Weyl spinors $\chi_{\alpha}^I $,
and two real scalars $\D$ and $\D^\prime $.
In four space-time dimensions an antisymmetric tensor only contains
one physical degree of freedom and is dual to a real scalar 
$a(x)$ via 
$\epsilon^{\mu\nu\rho\sigma}\partial_\mu b_{\nu\rho}
\sim \partial ^\sigma a$.
This duality can be elevated to a duality between entire
supermultiplets and one finds that 
the vector-tensor multiplet is dual to a vector multiplet
while 
the tensor multiplet and
the double tensor multiplet are both dual to a hyper multiplet.
Thus, the low energy effective theory can be described entirely
in terms of only the gravitational multiplet, vector- and
hypermultiplets. 
In particular, the moduli of  string compactifications 
appear either in vector- or hypermultiplets.  
Supersymmetry prohibits gauge neutral interactions 
between vector and
hypermultiplets and therefore the moduli space locally 
has to be a direct product  \cite{vanP}
\begin{equation}
{\cal M} = {\cal M}_H \times {\cal M}_V ,
\end{equation}
where ${\cal M}_H$ is the (quaternionic) moduli space parameterized by the
scalars of the hypermultiplets and ${\cal M}_V$ 
is the moduli space spanned
by the scalars in the vector multiplets. 

In the heterotic string the dilaton is  a member of a vector-tensor
multiplet or equivalently a  dual vector multiplet 
whereas in type IIA (IIB) 
the dilaton sits in a tensor (double tensor) multiplet 
or equivalently in the dual hypermultiplet.
This assignment of the dilaton together with the fact that
the dilaton organizes the string perturbation theory
immediately leads to two non-renormalization theorems. 
In type II compactifications the ${\cal M}_V$ component
of the moduli space cannot receive any perturbative or
non-perturbative corrections and the string tree level
result is exact. Conversely, in  heterotic
compactifications the ${\cal M}_H$ component 
suffers no quantum corrections.\cite{vector-tensor,stromingerBH}

Supersymmetry also dictates the local geometry of
these moduli spaces; 
${\cal M}_H$ is a quaternionic manifold\cite{BW}
while ${\cal M}_V$ is special K\"ahler.\cite{dWvP}
Because of its technical simplicity  most investigations 
so far focussed on  the special K\"ahler manifold
${\cal M}_V$.  
A special K\"ahler manifold is a K\"ahler manifold
with the metric $G_{i \bar{j}}$ expressed 
in terms of a K\"ahler potential $K$,
\begin{equation}
G_{i \bar{j}} = \frac{\partial}{\partial \phi^i}
\frac{\partial}{\partial \bar{\phi}^{\bar{j}}}
K\left( \phi, \bar{\phi}\right),
\end{equation}
where $\phi^i$ are the complex scalars in the vector multiplets. 
The term
`special' refers to the fact that the K\"ahler potential 
satisfies the additional constraint
\begin{equation}      \label{eq:kahlerpot}
K = -\log \left[ 2 \left( {\cal F}+\bar{{\cal F}}\right) -
\left( \phi^i -\bar{\phi}^{\bar{i}}\right)\left( {\cal F}_i - \bar{\cal F}_{\bar{i}}
\right)\right]\ , 
\qquad {\cal F}_i \equiv \frac{\partial{\cal F} }{\partial \phi^i}\ .
\end{equation}
That is, $K$ is determined by a single 
holomorphic function ${\cal F}(\phi)$ termed prepotential.
The same prepotential also determines all the gauge couplings
of the vector multiplets so that all couplings
of low energy effective
Lagrangian are expressed in terms of the holomorphic 
${\cal F}(\phi)$.\footnote{The precise expression
is of no concern here but can be found in 
the literature.\cite{vPrev}}

%
\subsection{The heterotic string compactified
on $K3 \times T^2$}
Compactifying a six-dimensional $(1,0)$ supergravity
on a two-torus $T^2$ results in $N=2$ supergravity 
in $d=4$ or equivalently  
the ten-dimensional heterotic string compactified 
on $K3\times T^2$ has $N=2$ supergravity in $d=4$. 
The massless perturbative spectrum
consists of the dilaton multiplet which is a
vector-tensor
or dual vector multiplet denoted by $S$.
In addition, there are two Abelian vector
multiplets $T$ and $U$ which contain the
 toroidal moduli of $T^2$ as well as
further  model-dependent
Abelian or possibly  non-Abelian 
vector multiplets.
The total perturbative gauge group is
\begin{equation}
G= G'\times U(1)_S\times U(1)_T\times U(1)_U\times U(1)_\gamma ,
\end{equation} 
where $G'$ refers to the additional
Abelian or non-Abelian part of the gauge group
and $U(1)\gamma$  corresponds to the graviphoton.
Furthermore, there are charged as well as
neutral hypermultiplets but their interaction 
will be of no concern here.
However, as we learned in the previous sections 
at special points
in the hypermultiplet moduli space
there might be a non-perturbative enhancement 
of this gauge group.

The prepotential for this class of vacua is found to
be \cite{vector-tensor,AFGNT}
\begin{equation}
{\cal F}= {\cal F}^{(0)}+{\cal F}^{(1)} + {\cal F}^{(NP)},
\end{equation}
where ${\cal F}^{(0)}$ and ${\cal F}^{(1)}$ denote
the tree level and one-loop contributions, respectively
 while 
${\cal F}^{(NP)}$ denotes the a priori unknown
non-perturbative corrections.
The absence of any perturbative contributions beyond
one-loop is a consequence of the 
$N=2$ non-renormalization theorem.\cite{kristin}

For all heterotic vacua 
the tree level contribution is found to be
\cite{fvp} 
\begin{equation} \label{eq:pre-het}
{\cal F}^{(0)}  = -S\left( TU - \phi^i\phi^i\right),
\end{equation}
where the $\phi^i$ are the vector multiplet moduli 
of the factor $G'$.
The one loop contribution ${\cal F}^{(1)}$ is 
model dependent but does not 
depend on the dilaton $S$. It  is strongly
constrained by the T-duality of the two-torus
$T^2$ and has been
computed for particular classes of 
heterotic vacua.\cite{vector-tensor,AFGNT,
MS}\cto\cite{gabriel4}


\subsection{Type II vacua
 compactified on Calabi-Yau threefolds}
Calabi-Yau threefolds are Calabi-Yau manifolds of complex dimension
three and holonomy group $SU(3)$.\cite{tex1,mirror}
For threefolds the hodge diamond is given by
\begin{equation} \label{eq:hodge-3}
\begin{array}{c c c c c c c}
           &  &              &  1 &            &  &          \\
      & &      0\! &  & \! 0 & &                         \\
  & 0\! & &h^{1,1} & &\! 0 & \\
  1\! & & h^{1,2}\! & & \! h^{1,2}& & \! 1\\
   & 0\! & & h^{1,1} & & \! 0& \\
 & & 0\! & &\! 0 & & \\
 & & & 1 & & &  
\end{array}\ ,
\end{equation}
where $h^{1,1}$ and $h^{1,2}$ are arbitrary.

It is believed that most (if not all) 
Calabi--Yau threefolds have an associated 
mirror manifold $\tilde{Y}$ with reversed 
Hodge numbers, i.e.~$h^{1,1}(\tilde{Y})=h^{1,2}(Y)$ 
and
$h^{1,2}(\tilde{Y})=h^{1,1}(Y)$.\cite{mirror}
As a consequence type IIA theory
compactified on $Y$ is perturbatively 
equivalent to type IIB theory compactified on the 
mirror $\tilde{Y}$. 

The non-trivial $(1,1)$ and $(1,2)$ forms on the
threefold correspond to massless modes in 
string theory. 
In particular, for type IIA one has $n_V = h^{1,1}$
and $n_H = h^{1,2} +1$ with 
the extra hypermultiplet being 
the type II dilaton.\footnote{In type IIB this assignment
is exactly reversed and one has instead 
$n_V = h^{1,2}$ and $n_H = h^{1,1} +1$.}
As we already argued in section~8.1  this fact
renders the type II vector multiplet
moduli space exact with no
perturbative or non-perturbative corrections.
The prepotential is entirely determined at the string 
tree level and in the limit of large Calabi--Yau
manifolds it obeys \cite{mirror}
\begin{equation} \label{eq:pre-2a}
{\cal F} = - \frac{i}{6} d_{\alpha\beta\gamma}t^\alpha t^\beta t^\gamma
+ {\rm worldsheet\;  instantons}\ ,
\end{equation}
where $t^\alpha, \alpha = 1,\ldots,h^{1,1}$ 
denote the scalar moduli of the 
vector multiplets and
$d_{\alpha \beta\gamma}$ are the intersection numbers.
The contribution of worldsheet instantons
vanishes in the large volume limit 
$t_\alpha \to \infty$.

The perturbative gauge group of 
type IIA compactifications
is the Abelian group
\begin{equation}
G = U(1)^{h^{1,1} +1}
\end{equation}
where the additional $U(1)$ corresponds 
to the graviphoton. 

%
\subsection{Heterotic -- type II duality}
It has been conjectured that 
heterotic strings compactified on $K3\times T^2$ 
and type II strings compactified on 
Calabi-Yau threefolds are quantum equivalent.\cite{kachru,fhsv} 
A necessary condition for this duality to be true
is the agreement of the two prepotentials.
This immediately implies that the heterotic dilaton $S$
is not mapped to the type II dilaton but rather
to a modulus of the Calabi--Yau threefold, say $t^S$.
Thus, the heterotic--type II duality is slightly 
different than the dualities
considered so far in that the strong coupling regime
is not mapped
to the weak coupling regime of the dual theory 
but rather to an other perturbatively accessible
region of the moduli space. 

The heterotic--type II duality has been supported
by a number of explicit pairs of string vacua 
where the type II prepotential 
and also the perturbative heterotic 
${\cal F}^{(0)}+{\cal F}^{(1)}$
is known.\cite{kachru}\cto\cite{mayr,aldazabal,berglund,sonne}
The matching of these functions
is a highly non-trivial check and furthermore
`predicts' the non-perturbative ${\cal F}^{(NP)}$
of the heterotic vacuum.
In fact there is a well-defined subclass
of Calabi-Yau threefolds known as 
$K3$-fibrations\footnote{Here the base is a ${\bf CP_1}$
and the fibre a $K3$.}
which do satisfy all necessary
requirements for being the dual of 
perturbative heterotic vacua.\cite{mayr,luy}   
In particular, this class of type II vacua 
naturally explains the perturbative bound on the size
of the heterotic gauge group and also captures
the possibility of non-perturbative enhancement
of $G$.

In $N=2$ supergravity there are also certain
higher derivate couplings which are governed by holomorphic
functions. These couplings are of the form
\begin{equation}
g_n^{-2}\left(\phi\right) R^2 F^{2n-2}\ ,
\end{equation}
where $R$ is the Riemann curvature and $F$ the field strength
of the graviphoton.
The heterotic--type II duality also requires
these couplings to coincide for a given dual pair.
Indeed this has been verified for specific examples
providing yet another non-trivial test
of the duality.\cite{vadim,anton,gabriel3}

\section{Conclusion}

It is always difficult to conclude a subject 
which is currently `in full swing'.
In fact after the presentation of this lecture 
many more exciting advances have taken place
which cannot be covered here.
All tests of duality have so far succeeded
in one way or another so that one is tempted
to believe that there is some truth in this story.

\section*{Acknowledgements} 
The work of S.F.~and J.L.~is supported by 
GIF -- the German Israeli foundation for
scientific research. S.F.\ would like to thank Debashis Ghoshal,
Gautam Sengupta, Stefan Schwager, Stefan Theisen and especially 
Kristin F\"orger
for valuable discussions.
J.L.~thanks Stefan Theisen for many helpful discussions on the topics presented in this lecture and for a critical reading of the manuscript.

\vskip 1cm
%

\end{document}

%% file: fig1.pstex_t
\begin{picture}(0,0)%
\epsfig{file=fig1.pstex}%
\end{picture}%
\setlength{\unitlength}{0.00050000in}%
\begingroup\makeatletter\ifx\SetFigFont\undefined
\def\x#1#2#3#4#5#6#7\relax{\def\x{#1#2#3#4#5#6}}%
\expandafter\x\fmtname xxxxxx\relax \def\y{splain}%
\ifx\x\y   
\gdef\SetFigFont#1#2#3{%
  \ifnum #1<17\tiny\else \ifnum #1<20\small\else
  \ifnum #1<24\normalsize\else \ifnum #1<29\large\else
  \ifnum #1<34\Large\else \ifnum #1<41\LARGE\else
     \huge\fi\fi\fi\fi\fi\fi
  \csname #3\endcsname}%
\else
\gdef\SetFigFont#1#2#3{\begingroup
  \count@#1\relax \ifnum 25<\count@\count@25\fi
  \def\x{\endgroup\@setsize\SetFigFont{#2pt}}%
  \expandafter\x
    \csname \romannumeral\the\count@ pt\expandafter\endcsname
    \csname @\romannumeral\the\count@ pt\endcsname
  \csname #3\endcsname}%
\fi
\fi\endgroup
\begin{picture}(3308,1516)(893,-969)
\put(4201,-361){\makebox(0,0)[lb]{\smash{\SetFigFont{14}{16.8}{rm}$\sim g_s$}}}
\end{picture}

%% file: fig2.pstex_t
\begin{picture}(0,0)%
\epsfig{file=fig2.pstex}%
\end{picture}%
\setlength{\unitlength}{0.00041700in}%
\begingroup\makeatletter\ifx\SetFigFont\undefined
\def\x#1#2#3#4#5#6#7\relax{\def\x{#1#2#3#4#5#6}}%
\expandafter\x\fmtname xxxxxx\relax \def\y{splain}%
\ifx\x\y   
\gdef\SetFigFont#1#2#3{%
  \ifnum #1<17\tiny\else \ifnum #1<20\small\else
  \ifnum #1<24\normalsize\else \ifnum #1<29\large\else
  \ifnum #1<34\Large\else \ifnum #1<41\LARGE\else
     \huge\fi\fi\fi\fi\fi\fi
  \csname #3\endcsname}%
\else
\gdef\SetFigFont#1#2#3{\begingroup
  \count@#1\relax \ifnum 25<\count@\count@25\fi
  \def\x{\endgroup\@setsize\SetFigFont{#2pt}}%
  \expandafter\x
    \csname \romannumeral\the\count@ pt\expandafter\endcsname
    \csname @\romannumeral\the\count@ pt\endcsname
  \csname #3\endcsname}%
\fi
\fi\endgroup
\begin{picture}(5408,1816)(1343,-3969)
\put(3751,-3136){\makebox(0,0)[lb]{\smash{\SetFigFont{6}{7.2}{rm}$+$}}}
\put(6751,-3136){\makebox(0,0)[lb]{\smash{\SetFigFont{6}{7.2}{rm}$+\ldots$}}}
\end{picture}

%% file: london.bbl
\begin{thebibliography}{150}
%
\bibitem{FILQ}A.~Font, L.E.~Iba\~nez, D.~L\"ust
 and F.~Quevedo,
{\it ``Strong--weak coupling duality and nonperturbative
effects in string theory''}, Phys.\ Lett.\ {\bf B249} (1990) 35.
%
\bibitem{rey} S.J.\ Rey, {\it ``The confining phase of superstrings and
axionic strings''}, Phys.\ Rev.\ {\bf D43} (1991) 526.
%
\bibitem{stromsoli} A.~Strominger,
{\it ``Heterotic solitons''}
Nucl.\ Phys.\ {\bf B343} (1990) 167.
%
\bibitem{ruiz1} A.~Dabholkar, G.~W.~Gibbons, J.~A.~Harvey 
and F.~Ruiz-Ruiz,
{\it ``Superstrings and solitons''}, 
Nucl.\ Phys.\ {\bf B340} (1990) 33.
%
\bibitem{dufflu1} M.J.\ Duff and J.X.\ Lu, 
{\it ``Elementary five-brane solutions
of $D=10$ supergravity''}, 
Nucl.\ Phys.\ {\bf B345} (1991) 141;
{\it ``Remarks on string/five-brane duality''}, 
Nucl.\ Phys.\ {\bf B354} (1991) 129.
%
\bibitem{ruiz2}G.~Horowitz and A.~Strominger, 
{\it ``Black strings and p-branes''}, 
Nucl.\ Phys.\ {\bf B360} (1991) 197.
%
\bibitem{sen} A.~Sen,
 {\it ``$SL(2,Z)$ duality and magnetically 
charged strings''},
Int.\ J.\ Mod.\ Phys.\ A8  (1993) 5079, hep-th/9302038;
%
{\it ``Dyon-monopole bound states, selfdual 
harmonic forms on the multi-monopole moduli space, 
and $SL(2,Z)$ invariance in string theory''}, 
Phys.\ Lett.\ {\bf B329} (1994) 217, hep-th/9402032.
%
\bibitem{schwarz-sen} J.~Schwarz and A.~Sen, 
{\it ``Duality symmetric actions''},
Nucl.\ Phys.\ {\bf B411} (1994) 35, hep-th/9304154;
%
{\it ``Duality symmetries of 4-D heterotic strings''}, 
Phys.\ Lett.\ {\bf B312} (1993) 105, hep-th//9305185.
%
\bibitem{DK} M.J.~Duff and R.R.~Khuri,
{\it ``Four-dimensional string/string duality''},
Nucl.\ Phys.\ {\bf B411} (1994) 473, hep-th/9305142.
%
\bibitem{duffme}M.J.~Duff and R.~Minasian, 
{\it ``Putting string/string duality to the test''}, 
Nucl.\ Phys.\ {\bf B436} (1995) 261, hep-th/9406198.
%
\bibitem{further0} C.~Vafa and E.~Witten, 
{\it ``A strong coupling test of S duality''}, 
Nucl.\ Phys.\ {\bf B431} (1994) 3, hep-th/9408074.
%
\bibitem{hull} C.M.~Hull and P.K.~Townsend, 
{\it ``Unity of superstring  dualities''},
Nucl.\ Phys.\ {\bf B438} (1995) 109, hep-th/9410167.
%
\bibitem{duff} M.J.~Duff,
{\it ``Strong/weak coupling duality from the 
dual string''},
Nucl.\ Phys.\ {\bf B442} (1995) 47, hep-th/9501030.
%
\bibitem{various} E.~Witten, 
{\it ``String theory dynamics in various dimensions''}, 
Nucl.\ Phys.\ {\bf B443} (1995) 85, hep-th/9503212.
%
\bibitem{sen-he} A.~Sen, 
{\it ``String-string duality conjecture in six dimensions
and charged solitonic strings''}, 
Nucl.\ Phys.\ {\bf B450} (1995) 103, 
hep-th/9504027.
%
\bibitem{hav-stro} J.A.~Harvey and A.~Strominger, 
{\it ``The heterotic string is a soliton''},
Nucl.\ Phys.\ {\bf B449} (1995) 456, hep-th/9504047.
%
\bibitem{aspe} P.S.~Aspinwall and D.R.~Morrison, 
{\it ``U duality and integral structures''}, 
Phys.\ Lett.\ {\bf B355} (1995) 141, hep-th/9505025.
%
\bibitem{kachru}S.~Kachru and C.~Vafa, 
{\it ``Exact results for $N=2$
compactifications of heterotic strings''}, 
Nucl.\ Phys.\ {\bf B450} (1995) 69, hep-th/9505105.
%
\bibitem{fhsv}S.~Ferrara, J.A.~Harvey, A.~Strominger and C.~Vafa,
{\it `` Second-quantized mirror symmetry''}, 
Phys.\ Lett.\ {\bf B361} (1995) 59, hep-th/9505162.
%
\bibitem{vadim} V.~Kaplunovsky, J.~Louis and S.~Theisen, 
{\it ``Aspects
of duality in $N=2$ string vacua''}, 
Phys.\ Lett.\ {\bf B357} (1995),71 hep-th/9506110.
%
\bibitem{mayr}A.~Klemm, W.~Lerche and P.~Mayr, 
{\it ``$K3$ fibrations and 
heterotic type II string duality''}, 
Phys.\ Lett.\ {\bf B357} (1995) 313,  hep-th/9506112.
%
\bibitem{type1a}A.~Dabholkar, 
{\it ``Ten dimensional heterotic string
as a soliton''}, 
Phys.\ Lett.\ {\bf B357} (1995) 307, hep-th/9506160.
%
\bibitem{type1b}C.M.~Hull, 
{\it ``String-string duality in ten-dimensions''}, 
Phys.\ Lett.\ {\bf B357} (1995) 545, hep-th/9506194.
%
\bibitem{VW}
C.~Vafa and E.~Witten,
{\it ``Dual string pairs with $N=1$ and $N=2$
supersymmetry in four-dimensions''}, hep-th/9507050.
%
\bibitem{anton}I.~Antoniadis, E.~Gava, K.S.~Narain and
T.R.~Taylor, 
{\it ``$N=2$ type II - heterotic duality and 
higher derivative F-terms''}, 
Nucl.\ Phys.\ {\bf B455} (1995) 109, hep-th/9507115.
%
\bibitem{wi-tension} E.~Witten, 
{\it ``Some comments on string dynamics''},
Contribution to STRINGS 95, Los Angeles, March 1995, hep-th/9507121.
%
\bibitem{SV}
A.~Sen and  C.~Vafa
{\it ``Dual pairs of type II string compactification''},
Nucl.\ Phys.\ {\bf B455} (1995) 165, hep-th/9508064.
%
\bibitem{aldazabal} G.~Aldazabal, A.~Font, L .E.~Iba\~{n}ez and
F.~Quevedo, 
{\it``Chains of $N=2,D=4$ heterotic type II duals''},
Nucl.\ Phys.\ {\bf B461} (1996) 85, hep-th/9510093;
{\it ``Heterotic/heterotic duality in $D=6,4$''}, 
Phys.\ Lett.\ {\bf B380} (1996) 33,  hep-th/9602097.
%
\bibitem{polch-witt}J.~Polchinski and E.~Witten, 
{\it ``Evidence for heterotic - type I duality''},
Nucl.\ Phys.\ {\bf B460} (1996) 525, hep-th/9510169.
%
\bibitem{horava1} P.~Ho\v{r}ava and E.~Witten, 
{\it ``Heterotic and type I string dynamics from eleven-dimensions''}, 
Nucl. Phys.\ {\bf B460} (1996),506, hep-th/9510209;
{\it ``Eleven-dimensional supergravity and a
manifold with boundary''}, hep-th/9603142.
%
\bibitem{gabriel2} G.~Lopes Cardoso, G.~Curio, 
D.~L\"ust, T.~Mohaupt and S.-J.~Rey, 
{\it ``BPS spectra and non-perturbative couplings
in $N=2,4$ supersymmetric string theories''}, 
Nucl.\ Phys.\ {\bf B464} (1996) 18, hep-th/9512129.
%
\bibitem{hul}C.M.~Hull, 
{\it ``String dynamics at strong coupling''},
Nucl.\ Phys.\ {\bf B468} (1996) 113, hep-th/9512181.
%
\bibitem{dasgupta} K.~Dasgupta and S.~Mukhi, 
{\it ``Orbifolds of M theory''},
Nucl.\ Phys.\ {\bf B465} (1996) 399, hep-th/9512196.
%
\bibitem{wittenm} E.~Witten, 
{\it ``Five-branes and M-theory on an orbifold''},
Nucl.\ Phys.\ {\bf B463} (1996) 383, hep-th/9512219.
%
\bibitem{duffmewi}M.J.~Duff, R.~Minasian and E.~Witten,
{\it ``Evidence for heterotic/heterotic duality''}
Nucl.\ Phys.\ {\bf B465} (1996) 413, hep-th/9601036.
%
\bibitem{vafas-f}C.~Vafa, {\it ``Evidence for F theory''},
Nucl.\ Phys.\ {\bf B469} (1996) 403, hep-th/9602022.
%
\bibitem{morva1}D.R.~Morrison and C.~Vafa, 
{\it ``Compactifications of F-theory
on Calabi-Yau threefolds - I''}, hep-th/9602111;
{\it ``Compactifications of 
F-theory on Calabi-Yau threefolds - II''}, 
hep-th/9603161.
%
\bibitem{AG} P.S.~Aspinwall and M.~Gross,
{\it ``Heterotic-heterotic string duality and multiple
$K3$ fibrations''},
Phys.\ Lett.\ {\bf B382} (1996) 81,  hep-th/9602118.
%
\bibitem{nati}N.~Seiberg and E.~Witten, 
{\it ``Comments on string dynamics in six dimensions''}, 
Nucl.\ Phys.\ {\bf B471} (1996) 121, hep-th/9603003.
%
\bibitem{AGK3} P.S.~Aspinwall and M.~Gross,
{\it ``The $SO(32)$ heterotic string on a $K3$ surface''},
Phys.\ Lett.\ {\bf B387} (1996) 735,  hep-th/9605131.
%
\bibitem{sen-f} A.~Sen, 
{\it ``F-theory and orientifolds''}, hep-th/9605150.
%
\bibitem{berglund} P.~Berglund, S.~Katz, A.~Klemm and P.~Mayr, 
{\it ``New
Higgs transitions between dual $N=2$ string models''}, hep-th/9605154.
%
\bibitem{bele}M.~Berkooz, R.~Leigh, J.~Polchinski,
J.~Schwarz, N.~Seiberg and E.~Witten, 
{\it ``Anomalies, Dualities, and Topology
of $D=6 \, N=1$ superstring vacua''}, 
Nucl.\ Phys.\ {\bf B475} (1996) 115, 
hep-th/9605184.
%
\bibitem{BIKMSV}
M.~Bershadsky, K.~Intriligator, S.~Kachru, D.R.~Morrison,
V.~Sadov and  C.~Vafa,
{\it ``Geometric singularities and enhanced gauge symmetries''},
hep-th/9605200.
%
\bibitem{sonne} J.~Louis, J.~Sonnenschein, S.~Theisen 
and S.~Yankielowicz, 
{\it ``Nonperturbative properties of heterotic
string vacua''}, hep-th/9606049.
%
\bibitem{gabriel3}B.~de Wit, G.~Lopes Cardoso,
D.~L\"ust, T.~Mohaupt and S.-J.~Rey, 
{\it ``Higher order gravitational 
couplings and modular forms in $N=2$, $D=4$ heterotic string 
compactifications''}, hep-th/9607184.
%
%
\bibitem{schwarzrev} J.~Schwarz,
{\it ``Lectures on superstring and M theory dualities''},
hep-th/9607021.
%
\bibitem{duffrev}
M.~Duff, 
{\it ``Supermembranes''}, hep-th/9611203.
%
\bibitem{d-brane4} J.~Polchinski, S.~Chaudhuri and C.V.~Johnson, 
{\it ``Notes on D-branes''}, hep-th/9602052;
J.~Polchinski, {\it ``TASI lectures on D-branes''},
hep-th/9611050.
%
\bibitem{senrev} A.~Sen,
{\it ``Unification of string dualities''},
hep-th/9609176.
%
\bibitem{townsendrev} P.K.~Townsend,
{\it ``Four lectures on M-theory''},
hep-th/9612121.
%
%
\bibitem{tex1} M.~Green, J.~Schwarz and E.~Witten, {\it ``Superstring theory''}, 
Vol.~1\&2, Cambridge University Press,  (1987).
%
\bibitem{tex2}D.~L\"ust and S.~ Theisen, 
{\it ``Lectures on string theory''},
Springer, (1989).
%
\bibitem{tex3}
J.~Polchinski, 
{\it ``What is String Theory''}, 
Les Houches lectures 1994, hep-th/94111028.
%
\bibitem{tex4}
M.~Peskin, 
{\it ``From the Planck Scale to the Weak Scale}'', 
ed.\ H.~Haber, World Scientific, (1987).
%
%
\bibitem{AAT}A.A.~Tseytlin, in
{\sl Superstrings '89}, 
ed.\ M.~Green, R.~Iengo, S.~Randjbar-Daemi, E.~Sezgin
and  A.~Strominger, 
World Scientific, (1990).\\
%
C.~Callan and L.~Thorlacius, 
in
{\sl Particles, Strings and Supernovae}, 
ed.\ A.~Jevicki and C.-I.~Tan, 
World Scientific, (1989).
%
\bibitem{ASNW} A. Schellekens and N. Warner,
{\it ``Anomalies and modular invariance in string 
theory''},
 Phys.\ Lett.\ 177B (1986) 317.
%
\bibitem{j-gang} A.~Giveon, M.~Porrati and E.~Rabinovici, 
{\it ``Target space duality in string theory''}, 
Phys.\ Rept.\ {\bf 244} (1994) 77, hep-th/9401139.
%
\bibitem{narain} K.S.~Narain, 
{\it ``New heterotic string theories in
uncompactified dimensions $< 10$''}, 
Phys.\ Lett.\ {\bf 169B} (1986) 41;\\
K.S.~Narain, M.H.~Samadi and E.~Witten,
{\it ``A note on toroidal compactification of 
heterotic string theory''}, 
Nucl.\ Phys.\ {\bf B279} (1987) 369.
%
\bibitem{gins} P.~Ginsparg, 
{\it ``Comment on  toroidal compactification
of heterotic superstrings''}, 
Phys.\ Rev.\ {\bf D35} (1987) 648.
%
\bibitem{berg}E.~Bergshoff, C.~Hull and T.~Ortin, 
{\it  ``Duality in the type II
superstring effective action''}, 
Nucl.\ Phys.\ {\bf B451} (1995) 547, 
hep-th/9504081.\\
%
A.~Kehagias, 
{\it ``Type IIA/IIB string duality for targets with
Abelian isometry''}, 
Phys.\ Lett.\ {\bf B377} (1996) 241, hep-th/ 9602059.
%
\bibitem{DHS}M.~Dine, P.~Huet and N.~Seiberg, 
{\it ``Large and small radius in string theory''},
Nucl.\ Phys.\ {\bf B322} (1989) 301.
%
\bibitem{d-brane1}
J.~Dai, R.G.~Leigh, J.~Polchinski, 
{\it ``New connections between string theories''}, 
Mod.\ Phys.\ Lett.\ {\bf A4} (1989) 2073.
%
\bibitem{d-brane2}P.~Ho\v{r}ava, 
{\it ``Strings on world sheet 
orbifolds''}, 
Nucl.\ Phys.\ {\bf B327} (1989) 461;
{\it ``Background duality of open
string models''}, 
Phys.\ Lett.\ {\bf B231} (1989) 251.
%
\bibitem{d-brane5}C.~Klim\v{c}\'{\i}k and P.~Severa, 
{\it ``Poisson Lie T 
duality: open strings and D-branes''}, 
Phys.\ Lett.\ {\bf B376} (1996) 82,
hep-th/9512124.
%
\bibitem{d-brane6}C.~Schmidhuber, 
{\it ``D-brane actions''},
Nucl.\ Phys.\ {\bf B467} (1996) 146, hep-th/9601003.
%
\bibitem{d-brane7}E.~Alvarez, J.L.F.~Barbon and  J.~Borlaf, 
{\it ``T--duality for open strings''} 
Nucl.\ Phys.\ {\bf B479} (1996) 218, hep-th/9603089.
%
\bibitem{d-brane8} H.~Dorn and H.-J.~Otto, 
{\it ``On T--duality
for open strings in general Abelian and non-Abelian gauge
field backgrounds''}, 
Phys.\ Lett.\ {\bf B381} (1996) 81, hep-th/9603186.
%
\bibitem{d-brane9}S.~F\"orste, A.~Kehagias and 
S.~Schwager, 
{\it ``Non-Abelian 
duality for open strings''}, 
Nucl.\ Phys.\ {\bf B478} (1996) 141, 
hep-th/9604013.
%
\bibitem{d-brane10} Y.~Lozano and J.~Borlaf, 
{\it ``Aspects
of T--duality in open strings''}, 
Nucl.\ Phys.\ {\bf B480} (1996) 239,
hep-th/9607051.
%
\bibitem{mo-ol} C.~Montonen and D.~Olive, 
{\it ``Magnetic monopoles
as gauge particles?''}, 
Phys.\ Lett.\ {\bf 72B} (1977) 117.
%
\bibitem{mo-ol-su1} E.~Witten and D.~Olive, 
{\it ``Supersymmetry algebras
that include topological charges''}, 
Phys.\ Lett.\ {\bf 78B} (1978) 97.
%
\bibitem{mo-ol-su2} H.~Osborn, 
{\it ``Topological charges
for $N=4$ supersymmetric gauge theories and monopoles of spin 1''},
Phys.\ Lett.\ {\bf 83B} (1979) 321.
%
\bibitem{mo-ol-su3} E.~Witten, 
{\it ``Dyons of charge $e\theta/2\pi$''},
Phys.\ Lett.\ {\bf 86B} (1979) 283.
%
\bibitem{harveyrev} For a review see, J.A.~Harvey
{\it ``Magnetic monopoles, duality and supersymmetry''},
hep-th/9603086.
%
\bibitem{we-ba} J.~Wess and J.~Bagger, 
{\it ``Supersymmetry and
Supergravity''}, 
Princeton University Press  (1983).
%
\bibitem{thooft} G.~'t Hooft, {\it ``Magnetic monopoles in
unified gauge theories''},
Nucl.\ Phys.\ {\bf B79} (1974) 276.
%
\bibitem{poly} A.M.~Polyakov,
{\it ``Particle spectrum in the quantum field theory''},
JETP Lett.\ {\bf 20} (1974) 194.
%
\bibitem{JZ}
B.~Julia and  A.~Zee,
{\it ``Poles with both magnetic and electric charges 
in nonabelian gauge theories''},
Phys.\ Rev.\ {\bf D11} (1975) 2227. 
%
\bibitem{dirac1} P.A.M.~Dirac, 
{\it ``Quantized singularities in the
electromagnetic field''}, 
Pro.\ R.\ Soc.\ {\bf A133} (1931) 60.
%
\bibitem{dirac2}D.~Zwanziger, 
{\it ``Exactly soluble non relativistic model
of particles with both electric and magnetic charges''}, 
Phys.\ Rev.\ {\bf 176} (1968) 1480;
{\it ``Quantum field theory of particles with both
electric and magnetic charges''}, 
Phys.\ Rev.\ {\bf 176} (1968) 1489.
%
\bibitem{dirac4}J.~Schwinger, 
{\it ``Magnetic charge and quantum 
field theory''}, 
Phys.\ Rev.\ {\bf 144} (1966) 1087;
{\it ``Sources and magnetic charge''}, 
Phys.\ Rev.\ {\bf 173} (1968) 1536.
%
\bibitem{bps1}M.K.~Prasad and C.M.~Sommerfield,
{\it ``An exact classical solution for the `t Hooft monopole and the
Julia-Zee condition''}, 
Phys.\ Rev.\ Lett.\ {\bf 35} (1975) 760.
%
\bibitem{bps2}E.B.~Bogomolny, 
{\it ``Stability of classical solutions''},
Sov.\ J.\ Nucl.\ Phys.\ {\bf 24} (1976) 449.
%
\bibitem{FSZ}
S.~Ferrara, C.A.~Savoy and B.~Zumino,
{\it ``General massive multiplets in extended
supersymmetry''},
Phys.\ Lett.\ {\bf 100B} (1981) 393.
%
\bibitem{further1}
J.~Gauntlett and J.A.~Harvey,
{\it ``S-duality and the dyon spectrum in 
$N=2$ super Yang--Mills theory''},
Nucl.\ Phys.\ {\bf 463} (1996) 287, hep-th/950815.
%
\bibitem{further2} M.~Porrati, 
{\it ``On the existence of states 
saturating the Bogomol'ny bound in $N=4$ supersymmetry''},
Phys.\ Lett.\ {\bf B377} (1996) 67, hep-th/9505187.
%
\bibitem{gimon} E.G.~Gimon and J.~Polchinski, 
{\it ``Consistency conditions
for orientifolds and D-manifolds''}, 
Phys.\ Rev.\ {\bf D54} (1996) 1667, 
hep-th/9601036.
%
\bibitem{callan}C.G.~Callan, J.A.~Harvey and A.~Strominger, 
{\it ``World-sheet approach to heterotic instantons and solitons''},
Nucl.\ Phys.\ {\bf B359} (1991) 611;
{\it ``World brane actions for string solitons''}, 
Nucl.\ Phys.\ {\bf B367} (1991) 60;
{\it ``Supersymmetric string solitons''}, 
hep-th/9112030.
%
\bibitem{duffPR} For a review see, 
M.J.~Duff, R.R.~Khuri and J.X.~Lu, 
{\it ``String solitons''},
Phys.\ Rept.\ {\bf 259} (1995) 213, hep-th/9412184. 
%
\bibitem{leigh} R.\ G.\ Leigh, 
{\it ``Dirac-Born-Infeld action from Dirichlet
$\sigma$-Model''}, 
Mod.\ Phys.\ Lett.\ {\bf A4}(1989)2767.
%
\bibitem{cremmer} E.~Cremmer and B.~Julia, 
{\it ``The $N=8$ supergravity 
theory. I. The Lagrangian''}, 
Phys.\ Lett.\ {\bf 80B} (1978) 48.
%
\bibitem{GZ} M.K.~Gaillard and B.~Zumino,
{\it ``Duality rotations for interacting fields''},
Nucl.\ Phys.\ {\bf B193} (1981) 221. 
%
\bibitem{JSchwarz} J.~Schwarz, 
{\it ``An $SL(2,Z)$ multiplet of type IIB
superstrings''}, 
Phys.\ Lett.\ {\bf B360} (1995) 13, hep-th/9508143  .
%
\bibitem{nahm} W.~Nahm, 
{\it ``Supersymmetries and their representations''},
Nucl.\ Phys.\ {\bf B135} (1978) 149.
%
\bibitem{cjs} E.~Cremmer, B.~Julia and J.~Scherk, 
{\it ``Supergravity theory in 11 dimensions''.} 
Phys.\ Lett.\ {\bf 76B} (1978) 409.
%
\bibitem{huq1} M.~Huq and M.~A.~Namazie, 
{\it ``Kaluza--Klein supergravity
in ten dimensions''}, 
Class.\ \& Quant.\ Grav.\ {\bf 2} (1985) 293.
%
\bibitem{huq2}F.~Giani and M.~Pernici, 
{\it ``$N=2$ supergravity in 
ten dimensions''}, 
Phys.\ Rev.\ {\bf D30} (1984) 325.
%
\bibitem{huq3}I.C.G.~Campbell and P.C.~West, 
{\it ``$N=2$ $D=10$ nonchiral
supergravity and its spontaneous compactification.''} 
Nucl.\ Phys.\ {\bf B243} (1984) 112.
%
\bibitem{gaume}  L.~Alvarez-Gaum\'{e} and E.~Witten, 
{\it ``Gravitational anomalies''}, 
Nucl.\ Phys.\ {\bf B234} (1984) 269.
%
\bibitem{green-schwarz} M.B.~Green and J.~Schwarz, 
{\it  ``Anomaly 
cancellations in supersymmetric $d=10$ gauge theory and
superstring theory''}, 
Phys.\ Lett.\ {\bf 149B} (1984) 117.
%
\bibitem{aspinwallrev} For a review see, P.~Aspinwall, 
{\it  ``$K3$ surfaces and string duality''}, 
hep-th/9611137.
%
\bibitem{NS} N.~Seiberg,
{\it  ``Observations on the moduli space of superconformal field
theories''},
Nucl.\ Phys.\ B303 (1988) 286. 
%
\bibitem{aspinwall} P.S.~Aspinwall and D.R.~Morrison, 
{\it ``String theory on $K3$ surfaces''}, hep-th/9404151.
%
\bibitem{townsend} P.K.~Townsend,
{\it ``A new anomaly free chiral supergravity theory
from compactification on $K3$''}, 
Phys.\ Lett.\ {\bf 139B} (1984) 283.
%
\bibitem{romans} L.~Romans
{\it ``Self-duality for interacting fields: covariant field
equations for six-dimensional chiral supergravities},
Nucl.\ Phys.\ {\bf B276} (1986) 71. 
%
\bibitem{seiwi1} N.~Seiberg and E.~Witten, 
{\it ``Electric magnetic duality, monopole
condensation and confinement in N=2 supersymmetric Yang-Mills 
theory.''},
Nucl.\ Phys.\ {\bf B426} (1994) 19, hep-th/9407087;
{\it ``Monopoles, duality and chiral symmetry 
breaking in N=2 supersymmetric QCD''}, 
Nucl.\ Phys.\ {\bf B431} (1994) 484,
hep-th/9408099.
%
\bibitem{aspinADE} P.S.~Aspinwall,
{\it ``Enhanced gauge symmetries and $K3$ surfaces''}, 
Phys.\ Lett.\ {\bf B357} (1995) 329, hep-th/9507012.
%
\bibitem{stromingerBH} A.~Strominger
{\it ``Massless black holes and conifolds in string theory''}
Nucl.\ Phys.\ {\bf B451} (1995),
hep-th/9504090.
%
\bibitem{gswest} M.B.~Green, J.~Schwarz and P.~C.~West, 
{\it ``Anomaly-free
chiral theories in six dimensions''}, 
Nucl.\ Phys.\ {\bf B254} (1985) 327.
%
\bibitem{erler} J.~Erler, 
{\it ``Anomaly cancellation in six dimensions''},
J.\ Math.\ Phys.\ {\bf 35} (1994) 1819, hep-th/9304104.
%
\bibitem{schwarzanom} J.~Schwarz, 
{\it ``Anomaly-free supersymmetric models in six dimensions''},
Phys.\ Lett.\ {\bf B371} (1996) 223, 
hep-th/9512053.
%
\bibitem{sagnotti} A.~Sagnotti, {\it ``A note on the Green-Schwarz mechanism
in open - string theories''}, Phys.\ Lett.\ {\bf B294} (1992) 196,
hep-th/9210127
%
\bibitem{lue}M.J.~Duff, H.~L\"u and C.N.~Pope, 
{\it ``Heterotic phase
transitions and singularities of the gauge dyonic string''}, 
Phys.\ Lett.\ {\bf B378} (1996) 101, hep-th/9603037.
%
\bibitem{schmal}E.~Witten, 
{\it ``Small instantons in string theory.''},
Nucl.\ Phys.\ {\bf B460} (1996) 541, hep-th/9511030.
%
\bibitem{hanany}O.J.~Ganor and A.~Hanany, 
{\it ``Small $E_8$ instantons
and tensionless non critical strings''}, 
Nucl.\ Phys.\ {\bf B474} (1996) 122,
hep-th/9602120.
%
\bibitem{stringy-cosmic} B.R.~Greene, A.~Shapere, C.~Vafa and S.-T.~Yau,
{\it ``Stringy cosmic strings and noncompact Calabi-Yau Manifolds''},
Nucl.\ Phys.\ {\bf B337} (1990) 1.
%
%
%
\bibitem{kristin} J.~Louis and K.~F\"orger, 
{\it ``Holomorphic couplings in 
string theory''}, hep-th/9611184.
%
\bibitem{vanP} B.~De Wit, P.G.~Lauwers and A.~van Proeyen, 
{\it ``Lagrangians of $N=2$ super-gravity matter systems''}, 
Nucl.\ Phys.\ {\bf B255} (1985) 569.
%
%
\bibitem{vector-tensor} B.~de Wit, V.~Kaplunovsky, J.~Louis and D.~L\"ust,
{\it ``Perturbative couplings of vector multiplets in $N=2$ heterotic 
string vacua''}, Nucl.\ Phys.\ {\bf B451} (1995) 53, 
hep-th/9504006.
%
\bibitem{BW} J.~Bagger and E.~Witten
{\it ``Matter couplings in $N=2$ supergravity''}, 
Nucl.\ Phys.\ {\bf B222} (1983) 1.
%
\bibitem{dWvP} B.~De Wit and A.~van Proeyen, 
{\it ``Potentials and symmetries of general gauged
$N=2$ supergravity - Yang-Mills models''}, 
Nucl.\ Phys.\ {\bf B245} (1984) 89.
%
\bibitem{vPrev} For a recent review on special
geometry see for example,
A.~van Proeyen, 
{\it ``Vector multiplets in $N=2$ supersymmetry
and its associated moduli spaces''}, hep-th/9512139.
%
\bibitem{AFGNT}I.~Antoniadis, S.~Ferrara, E.~Gava, K.S.~Narain and
T.R.~Taylor, 
{\it ``Perturbative prepotential and monodromies
in $N=2$ heterotic superstring''}, 
Nucl.\ Phys.\ {\bf B447} (1995) 35, hep-th/9504034.
%
\bibitem{fvp}S.~Ferrara and A.~van Proeyen, 
{\it ``A theorem on $N=2$
special K\"ahler product manifolds''}, 
Class.\ Quant.\ Grav.\ {\bf 6} (1989) L243.
%
\bibitem{MS} P.~Mayr and  S.~Stieberger 
{\it ``Moduli dependence of one loop
gauge couplings in $(0,2)$ compactifications''},
Phys.\ Lett.\ {\bf B355} (1995) 107, hep-th/9504129.
%
\bibitem{HM} J.A.~Harvey and G.~Moore,
{\it ``Algebras, BPS states, and strings''}
Nucl.\ Phys.\ {\bf B463} (1996) 315, hep-th/9510182.
%
\bibitem{HennM} M.~Henningson and G.~Moore
{\it ``Threshold corrections in $K3\times T^2$
heterotic string compactifications''},
hep-th/9608145.
%
\bibitem{gabriel4} G.~Lopes Cardoso, G.~Curio 
and D.~L\"ust,
{\it ``Perturbative couplings and modular forms in 
$N=2$ string models with a Wilson line''}, 
hep-th/9608154.
%
\bibitem{mirror} For a review see for example,
S.~Hosono, A.~Klemm and S.~Theisen,
{\it ``Lectures on mirror symmetry''},
hep-th/9403096.
%
\bibitem{luy}P.S.~Aspinwall and J.~Louis, 
{\it ``On the ubiquity of $K3$
fibrations in string duality''}, 
Phys.\ Lett.\ {\bf B369} (1996) 233, hep-th/9510234.
%
\end{thebibliography}
